\documentclass[preprint2]{aastex631}
\usepackage{graphicx}

\begin{document}

\title{Limits on the 21~cm power spectrum at $z=6.5-7.0$ from MWA observations}

\author[0000-0002-5445-6586]{C.~D.~Nunhokee}
\affiliation{International Centre for Radio Astronomy Research (ICRAR), Curtin University, Bentley, WA, Australia}
\affiliation{ARC Centre of Excellence for All Sky Astrophysics in 3 Dimensions (ASTRO 3D), Bentley, Australia} 
\affiliation{Curtin Institute of Radio Astronomy, GPO Box U1987, Perth, WA 6845, Australia}

\author[0000-0003-2159-0911]{D.~Null}
\affiliation{International Centre for Radio Astronomy Research (ICRAR), Curtin University, Bentley, WA, Australia}
\affiliation{Australian SKA Regional Centre (AusSRC), Curtin University, Bentley, WA, Australia}
\affiliation{ARC Centre of Excellence for All Sky Astrophysics in 3 Dimensions (ASTRO 3D), Bentley, Australia} 
\affiliation{Curtin Institute of Radio Astronomy, GPO Box U1987, Perth, WA 6845, Australia}

\author[0000-0001-6324-1766]{C.~M~Trott}
\affiliation{International Centre for Radio Astronomy Research (ICRAR), Curtin University, Bentley, WA, Australia}
\affiliation{ARC Centre of Excellence for All Sky Astrophysics in 3 Dimensions (ASTRO 3D), Bentley, Australia} 
\affiliation{Curtin Institute of Radio Astronomy, GPO Box U1987, Perth, WA 6845, Australia}

\author[0000-0003-2064-6979]{N.~Barry}
\affiliation{ARC Centre of Excellence for All Sky Astrophysics in 3 Dimensions (ASTRO 3D), Bentley, Australia} 
\affiliation{School of Physics, The University of New South Wales, Sydney, NSW 2052, Australia}

\author[0000-0002-4314-1810]{Y.~Qin}
\affiliation{Research School of Astronomy and Astrophysics, Australian National University, Canberra, ACT 2611, Australia}
\affiliation{ARC Centre of Excellence for All Sky Astrophysics in 3 Dimensions (ASTRO 3D), Bentley, Australia} 

\author{R.~B.~Wayth}
\affiliation{International Centre for Radio Astronomy Research (ICRAR), Curtin University, Bentley, WA, Australia}
\affiliation{ARC Centre of Excellence for All Sky Astrophysics in 3 Dimensions (ASTRO 3D), Bentley, Australia} 
\affiliation{Curtin Institute of Radio Astronomy, GPO Box U1987, Perth, WA 6845, Australia}

\author{J.~L.~B.~Line}
\affiliation{International Centre for Radio Astronomy Research (ICRAR), Curtin University, Bentley, WA, Australia}
\affiliation{ARC Centre of Excellence for All Sky Astrophysics in 3 Dimensions (ASTRO 3D), Bentley, Australia} 
\affiliation{Curtin Institute of Radio Astronomy, GPO Box U1987, Perth, WA 6845, Australia}

\author{C.~H.~Jordan}
\affiliation{International Centre for Radio Astronomy Research (ICRAR), Curtin University, Bentley, WA, Australia}
\affiliation{ARC Centre of Excellence for All Sky Astrophysics in 3 Dimensions (ASTRO 3D), Bentley, Australia} 
\affiliation{Curtin Institute of Radio Astronomy, GPO Box U1987, Perth, WA 6845, Australia}

\author{B.~Pindor}
\affiliation{School of Physics, The University of Melbourne, Parkville Vic 3010, Australia} 
\affiliation{ARC Centre of Excellence for All Sky Astrophysics in 3 Dimensions (ASTRO 3D), Bentley, Australia} 

\author{J.~H.~Cook}
\affiliation{International Centre for Radio Astronomy Research (ICRAR), Curtin University, Bentley, WA, Australia}
\affiliation{ARC Centre of Excellence for All Sky Astrophysics in 3 Dimensions (ASTRO 3D), Bentley, Australia} 
\affiliation{Curtin Institute of Radio Astronomy, GPO Box U1987, Perth, WA 6845, Australia}


\author{J.~Bowman}
\affiliation{School of Earth and Space Exploration, Arizona State University, AZ, USA}

\author[0000-0003-1130-6390]{A.~Chokshi}
\affiliation{The University of Melbourne, School of Physics, Parkville, VIC 3010, Australia}
\affiliation{ARC Centre of Excellence for All Sky Astrophysics in 3 Dimensions (ASTRO 3D), Bentley, Australia}

\author{J.~Ducharme}
\affiliation{Brown University, Providence RI, USA}

\author{K.~Elder}
\affiliation{School of Earth and Space Exploration, Arizona State University, AZ, USA}

\author{Q.~Guo}
\affiliation{Shanghai Astronomical Observatory, Chinese Academy of Sciences, 80 Nandan Road, Shanghai 200030, China}

\author{B.~Hazelton}
\affiliation{Department of Physics, University of Washington, WA, USA}

\author{W.~Hidayat}
\affiliation{Kumamoto University, Japan}

\author{T.~Ito}
\affiliation{Kumamoto University, Japan}

\author{D.~Jacobs}
\affiliation{School of Earth and Space Exploration, Arizona State University, AZ, USA}

\author{E.~Jong}
\affiliation{International Centre for Radio Astronomy Research (ICRAR), Curtin University, Bentley, WA, Australia}
\affiliation{ARC Centre of Excellence for All Sky Astrophysics in 3 Dimensions (ASTRO 3D), Bentley, Australia} 
\affiliation{Curtin Institute of Radio Astronomy, GPO Box U1987, Perth, WA 6845, Australia}

\author{M.~Kolopanis}
\affiliation{School of Earth and Space Exploration, Arizona State University, AZ, USA}

\author{T.~Kunicki}
\affiliation{Brown University, Providence RI, USA}

\author{E.~Lilleskov}
\affiliation{Department of Physics, University of Washington, WA, USA}

\author{M.~F.~Morales}
\affiliation{Department of Physics, University of Washington, WA, USA}

\author{J.~C.~Pober}
\affiliation{Department of Physics, Brown University, Providence RI, USA}

\author{A.~Selvaraj}
\affiliation{International Centre for Radio Astronomy Research (ICRAR), Curtin University, Bentley, WA, Australia}
\affiliation{ARC Centre of Excellence for All Sky Astrophysics in 3 Dimensions (ASTRO 3D), Bentley, Australia} 
\affiliation{Curtin Institute of Radio Astronomy, GPO Box U1987, Perth, WA 6845, Australia}

\author{R.~Shi}
\affiliation{Brown University, Providence RI, USA}
\affiliation{Astronomy Department, The University of Florida, FL, USA} 

\author{K.~Takahashi}
\affiliation{Kumamoto University, Japan}

\author{S.~J.~Tingay}
\affiliation{International Centre for Radio Astronomy Research (ICRAR), Curtin University, Bentley, WA, Australia}
\affiliation{Curtin Institute of Radio Astronomy, GPO Box U1987, Perth, WA 6845, Australia}

\author{R.~L.~Webster}
\affiliation{School of Physics, The University of Melbourne, Australia}
\affiliation{ARC Centre of Excellence for All Sky Astrophysics in 3 Dimensions (ASTRO 3D), Bentley, Australia} 

\author{S.~Yoshiura}
\affiliation{Mizusawa VLBI Observatory, National Astronomical Observatory of Japan, 2-21-1 Osawa, Mitaka, Tokyo 181-8588, Japan}

\author{Q.~Zheng}
\affiliation{Shanghai Astronomical Observatory, Chinese Academy of Sciences, 80 Nandan Road, Shanghai 200030, China}





\begin{abstract}
This paper presents the spherically-averaged 21~cm power spectrum derived from Epoch of Reionization (EoR) observations conducted with the Murchison Widefield Array (MWA). The analysis uses EoR0-field data, centered at (RA$=0h$, DEC$=-27^{\circ}$), collected between 2013 and 2023. Building on the improved methodology described in \citet{Trott2020}, we incorporate additional data quality control techniques introduced in \citet{Nunhokee2024}. We report the lowest power level limits on the EoR power spectrum at redshifts $z=6.5$, $z=6.8$, and $z=7.0$. These power levels, measured in the East--West polarization, are $(30.2)^2$ mK$^2$  at $k=0.18\, h$ Mpc$^{-1}$, $(31.2)^2$ mK$^2$ at $k=0.18\, h$ Mpc$^{-1}$ and $(39.1)^2$ mK$^2$ at $k=0.21\, h$ Mpc$^{-1}$ respectively. The total integration time amounts to $268$~hours. These results represent the deepest upper limits achieved by the MWA to date and provide the first evidence of heated intergalactic medium (IGM) at redshifts $z=6.5$ to $7.0$. 

\end{abstract}



\section{Introduction} 
\label{sec:intro}
The study of primordial hydrogen in the early Universe provides a deeper insight into the properties of the first galaxies, the physics of mini-quasars, the formation of very metal-poor stars, and the understanding of cosmic structure formation. Hydrogen was produced during Big Bang nucleosynthesis \citep{Kamionkowski2007}. It remained neutral until ultraviolet radiation from early massive stars or active galactic nuclei (AGN) escaped their host galaxies. This radiation began ionizing the surrounding IGM, initiating a period known as the Epoch of Reionization (EoR; \citealt{Zaroubi2013}). 

The EoR represents a pivotal stage in the Universe's evolution, and various probes are used to study this epoch. These include observations of the Gunn-Peterson trough in the spectra of distant quasars (e.g.~\citealt{Fan2006, Becker2015, Mortlock2016, Zhu2022}), the Lyman-$\alpha$ forest (e.g.~\citealt{Stark2011, Dijkstra2014, Pentericci2014, Bosman2022}), large-scale Cosmic Microwave Background (CMB) anisotropies (e.g.~\citealt{Planck2020}) and  fluctuations in the infrared extragalactic background light \citep{Olivier2014}. Observations of the 21~cm hyperfine line from neutral hydrogen are considered the most promising tool, as they directly probe the IGM (e.g.~\citealt{Barkana2001, Furnaletto2006}).

Measurements of the weak 21~cm signal are hindered by bright foregrounds, which are several orders of magnitude brighter, yielding systematic contamination that includes errors due to inaccurate calibration and spectral structures from radio galaxies and Galactic emission \citep{Jelic2008}. Efforts have been invested for the past few years into enhancing the EoR data processing through characterization of the foregrounds (e.g.~\citealt{Datta2010, Trott2012, Vedantham2012, Chapman2016, Thyagarajan2015a, Thyagarajan2015b, Eastwood2018, Mertens2018, Barry2024}), the calibration model (e.g.~\citealt{Offringa2015, Barry2016, Trott2016, Patil2017, Procopio2017, Dillon2018, Kern2019, Orosz2019, Lynch2021}), the instrument model (e.g.~\citealt{deLera2017, Trott2017, Li2018, Nunhokee2020}), the power spectrum estimation (e.g.~\citealt{Parsons2010, Parsons2012, Liu2014, Choudhari2017, Barry2019a, Kern2020}), RFI (e.g.~\citealt{Offringa2010, Wilensky2019, Kerrigan2019, Wilensky2023}), polarization leakage (e.g.~\citealt{Moore2017, Nunhokee2017}) and the effect of the ionosphere (e.g.~\citealt{Mevius2016, Jordan2017, Trott2018}).

Current low-frequency experiments actively geared towards measuring the 21~cm hydrogen line through statistical fluctuations of the IGM include the Hydrogen Epoch of Reionization Array (HERA; \citealt{DeBoer2017, Berkhout2024}), the Low-Frequency Array (LOFAR; \citealt{vanHaarlem2013}), and the Murchison Widefield Array (MWA; \citealt{Tingay2013, Wayth2018}). These experiments operate between redshifts of 6 to 13, while other instruments, such as the Long Wavelength Array (LWA; \citealt{Ellingson2009}) and the New Extension in Nan\c{c}ay Upgrading LOFAR (NenuFAR; \citealt{Munshi2024}), focus on statistical measurements at higher redshifts.

All the aforementioned experiments have reported upper limits on the 21~cm power spectrum. The best 21~cm upper limits of $\Delta^2 \leq (21.4)^2 \,\textrm{mK}^2$ at $k=0.34 \, h$ Mpc$^{-1}$ and $\Delta^2 \leq (59.1)^2\,\textrm{mK}^2$ at $k=0.36 \, h$ Mpc$^{-1}$ at redshifts of 7.9 and 10.4, respectively, have been reported by HERA \citep{HERA2023}. The latest findings indicate that heating of the intergalactic medium above the adiabatic cooling limit must have occurred by at least $z=10.4$, effectively ruling out a wide range of ``cold--reionization" models \citep{HERACollaboration2022A, HERA2023}.

Given that MWA EoR observations are more sensitive at lower redshifts, we focus on deriving the deepest upper limits on 21~cm power spectrum at $z=6.5$, $6.8$ and $7.0$, using a decade's worth of data. This work aims to place new constraints on, or further extend the exclusion of ``cold--reionization" models to lower redshifts.


We present the 21~cm experimental set up in Section~\ref{sec:observation} followed by the breakdown of the various stages involved in data processing in Section~\ref{sec:methodology}. The results are presented in Section~\ref{sec:results} and validated in Section \ref{sec:validation}. The astrophysics inference from our results is conducted in Section~\ref{sec:inference}. We discuss the findings and future work in Section~\ref{sec:discussion} and conclude in Section~\ref{sec:conclusion}.

\section{EoR Observing Program}
\label{sec:observation}
In this work, we used observations from the MWA, a low-frequency interferometric radio telescope located at \textit{Inyarrimanha Ilgari Bundara}, the Commonwealth Scientific and Industrial Research Organisation (CSIRO) Murchison Radio-astronomy Observatory. The instrument is situated in the mid-west of Western Australia, approximately 300 kilometers inland from the coastal town of Geraldton, due to its radio-quiet nature \citep{Offringa2015}.

\begin{table}[ht]
\caption{Altitudes and azimuths of telescope beam pointings, with zenith denoted by pointing 0 and -3 is three pointings before zenith, and 3 is three pointings after zenith.}
\label{tab:telescope_pointing}
\hspace{-1cm}
    \begin{tabular}{|c|c|c|}
    \hline
        East--West Pointing & Altitude $(^\circ)$ & Azimuth $(^\circ)$\\
        \hline
         Pointing -3& $69.2^{\circ}$ & $90^{\circ}$\\
         Pointing -2& $76.3^{\circ}$ & $90^{\circ}$\\
         Pointing -1& $83.2^{\circ}$ & $90^{\circ}$\\
         Pointing 0& $90^{\circ}$ & $0^{\circ}$\\
         Pointing 1& $83.2^{\circ}$ & $270^{\circ}$\\
         Pointing 2& $76.3^{\circ}$ & $270^{\circ}$\\
         Pointing 3& $69.2^{\circ}$ & $270^{\circ}$\\
         \hline
    \end{tabular}
\end{table} 
Data were captured using the Phase I and Phase II compact configurations across four polarizations: East--West (EW), North--South (NS), East--West/North--South (EW--NS) and North--South/East--West (NS--EW) (for detailed information, see \citealt{Tingay2013, Randall2018}). They spanned a frequency range of 167 to 197 MHz ($7.56 < z < 6.25$) from the target field EoR0, centered at (RA$=0h$, DEC$=-27^{\circ}$). The telescope observes using coarse pointing centers, spaced at intervals of $6.8^{\circ}$, to maintain a consistent primary beam response by aligning with fixed analogue delay settings \citep{Tingay2013}. Phase tracking remains constant on the observation field, EoR0, for all pointings.


Most observations were recorded at a resolution of 40 kHz with a 2 second cadence, although some of the observations used a finer cadence of 0.5 seconds. This resulted in raw file sizes ranging from 5 to 11~GB per 2 minute observation, with variations also attributed to later observations incorporating baselines from 145 tiles. Approximately $10\%$ of the observations were recorded at higher resolutions, producing raw file sizes of up to $247$ GB per observation. In total, 260.9~TB of raw data were processed.


\begin{table}
\centering
\caption{MWA Observing Parameters}
\label{tab:observation_parameters}
\scalebox{0.8}{
\begin{tabular}{ccc}
\hline \hline\\
Parameter & Phase I & Phase II Compact\\
 \hline
Field of View & 24.6$^{\circ}$ & 24.6$^{\circ}$\\
Maximum baseline &  2864~m & 749~m  \\
Angular resolution & $\sim$2~arcmin & $\sim$9~arcmin\\
Spectral resolution & 40~kHz & 40~kHz\\
Integration time & 2~s & 2~s\\
Observing Length & 112~s & 120~s\\
Observations & 9655 & 9970\\
\hline \hline
\end{tabular}
}
\end{table}

\begin{figure*}
\centering
\includegraphics[width=0.9\linewidth]{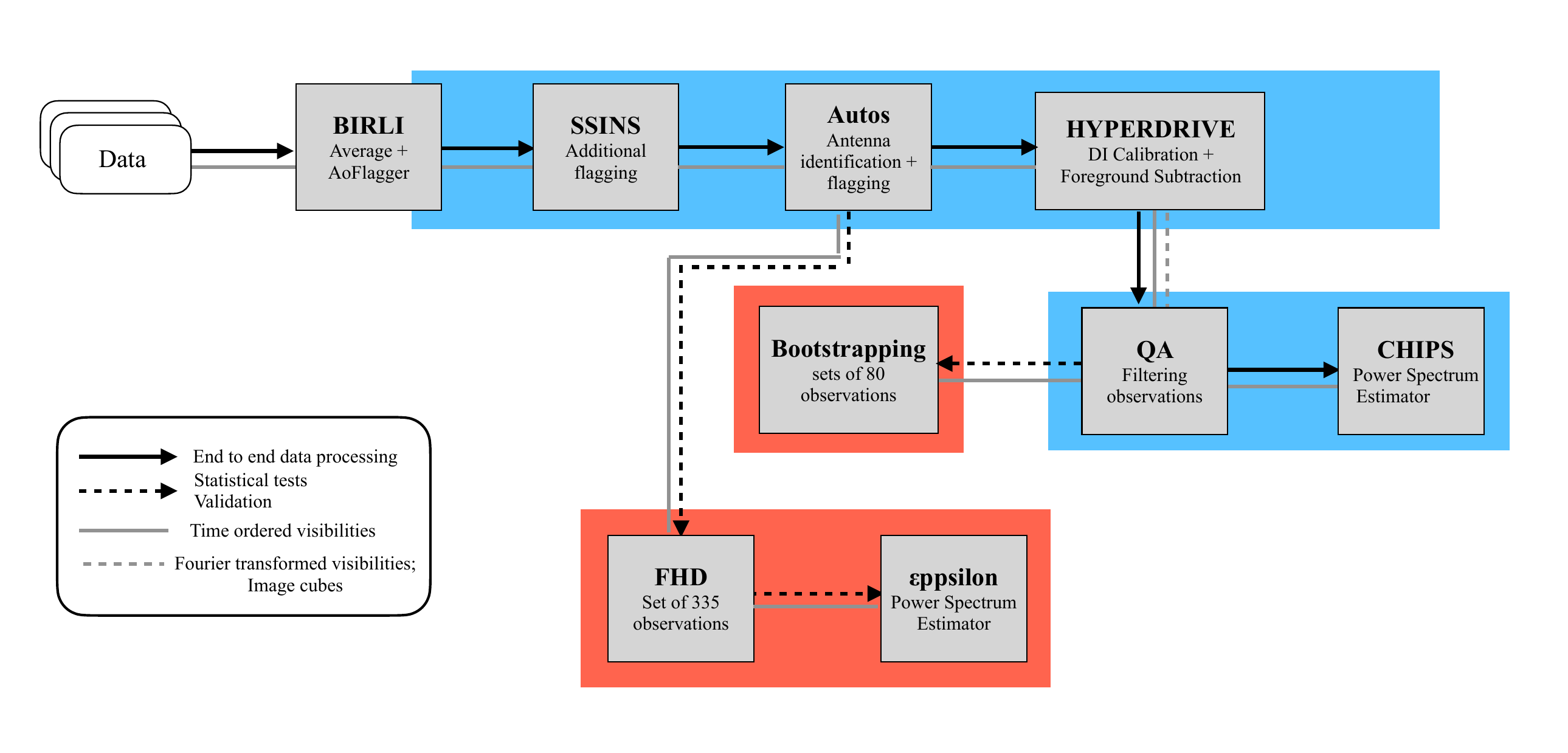}
\caption{The data processing pipeline, highlighted in blue, processes data end-to-end, starting with observations detailed in Section~\ref{sec:observation}. It outlines the various processes involved, specifies the tools used at each step, and illustrates the statistical and validation tests, enclosed in red, applied in this work.\\}
\label{fig:eor_pipeline}
\end{figure*}

\section{Methodology}
\label{sec:methodology}
This paper adopts a similar data processing and systematic mitigation strategies as \citet{Nunhokee2024}. We made a few improvements to the existing methodology which will be discussed in the following section. A schematic illustration of the processes and relevant tools used throughout the processing pipeline is presented in Figure~\ref{fig:eor_pipeline}. The pipeline is implemented in Nextflow.\footnote{\url{https://github.com/MWATelescope/MWAEoR-Pipeline}} 


\subsection{Flagging and Pre-processing}
\label{sec:flagging}
 We flag the data in several stages during preprocessing using Birli\footnote{\url{https://github.com/MWATelescope/Birli}}. Broadly, we flag 1) visibilities that are known to contain untrustworthy data, 2) Radio Frequency Interference (RFI) using AOFlagger \citep{Offringa2010} and the Sky-Subtracted Incoherent Noise Spectra (SSINS) \citep{Wilensky2019}. The MWA correlator initially performs channelization into 24 coarse channels, each 1.28~MHz wide, followed by fine channelization at 10~kHz resolution. However, for EoR observations, the effective spectral resolution is 40~kHz. Due to coarse sampling of the Fourier transform, the fine polyphase filterbank exhibits poor performance at the center and edges of each coarse channel. Consequently, we flag visibilities at both the center and edges of every coarse channel. In addition, visibilities at the beginning and end of each observation, as well as those identified as corrupted by the MWA Monitor and Control system, are also flagged.
Next, Birli uses AOFlagger to identify and flag RFI at the highest possible resolution before averaging the data to a frequency resolution of 40~kHz and a time resolution of 2~seconds. However, as noted by \citet{Wilensky2019}, AOFlagger's algorithm, which processes one baseline at a time, lacks sensitivity to RFI fainter than the thermal noise on a single baseline. To address this limitation, we supplement AOFlagger with SSINS algorithm on the averaged visibilities. We then perform a $z$-score analysis on auto-correlations to identify malfunctioning antennas, as detailed in \citet{Nunhokee2024}.

\subsection{Calibration}
\label{sec:calibration}
The visibilities are calibrated on a per-channel basis for direction-independent antenna gains using Hyperdrive\footnote{\url{https://mwatelescope.github.io/mwa_hyperdrive/index.html}}. This tool employs the MWA traditional sky-based calibration approach, comparing measured data with expected data from a known sky model and applying the necessary corrections \citep{Mitchell2008, Yattawata2008, Kazemi2011}. The calibration derives corrections for all four polarizations: EW, NS, EW-NS, and NS-EW.

As our initial calibration model, we combine data from the GLEAM survey \citep{Wayth2015, HurleyWalker2017} and the LoBES catalogue \citep{Lynch2021}, modeling the spectral energy distribution of sources as either a power-law or a curved power-law. We also implement several refinements to address issues identified in previous studies \citep{Procopio2017, Line2020, Cook2022}. These improvements include correcting mis-modeled sources highlighted by \citet{Procopio2017}, applying a shapelet model for Fornax A \citep{Line2020}, and incorporating models for Centaurus A and Galactic supernova remnants \citep{Cook2022}.

At each coarse frequency channel, we attenuate the sources' intrinsic flux densities using the Full Embedded Element (FEE) primary beam model generated with Hyperbeam \citep{Sokolowski2017}, and incorporated in Hyperdrive. From this, we select the brightest 8,000 sources for our final calibration model, using $70.6\%$ of the total flux densities of the source catalogue (excluding $\sim$1000~Jy worth of sources). 
We derive the direction-independent gain solutions from visibilities with baselines greater than 30 wavelengths, as short baselines are known to be dominated by diffuse emission (e.g.\citealt{Patil2017, Ruby2022}). 
While \citet{Patil2017} identified that baselines shorter than 50 wavelengths are prone to Galactic emission, adopting a 50 wavelengths cut-off in our case would discard approximately $32\%$ of baselines at 150~MHz for the Phase II compact configuration, limiting the improvements achievable in the calibration solutions \citep{Trott2016}.

To ensure the solutions converge optimally, we configure the underlying least squares minimization algorithm to perform up to 300 iterations. Increasing the iteration count from the Hyperdrive nominal default of 50 yields slight improvements in the calibration solutions. We then carefully inspect the resulting calibration solutions for amplitude deviations across antennas, rapidly migrating phases, and other relevant anomalies in each observation \citep{Nunhokee2024}. To identify outliers, we evaluate the root mean square of the amplitudes across frequencies and compare them among antennas. Additionally, we apply linear fits to the phases to detect any significant discrepancies.

\begin{figure}[h]
\centering
\includegraphics[width=1.0\linewidth]{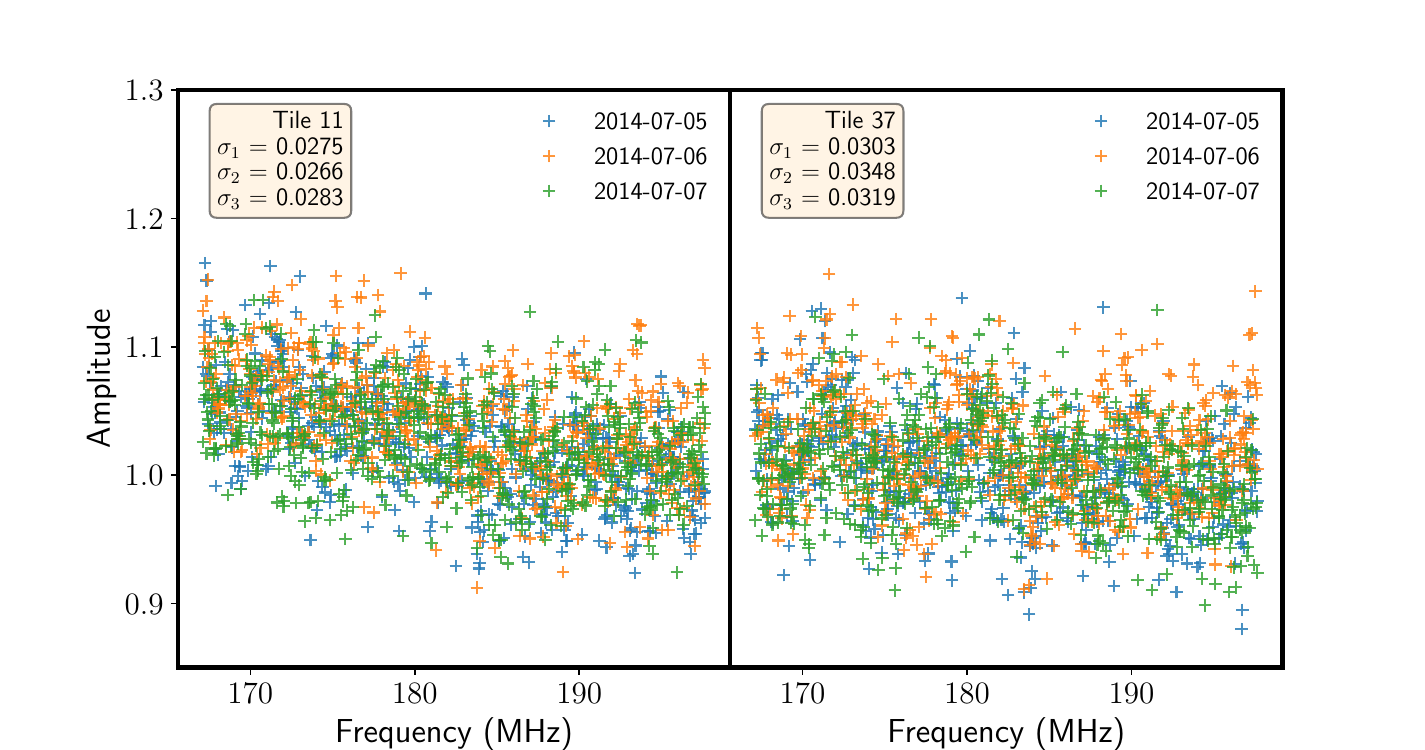}
\caption{The normalized amplitude of direction-independent antenna gains are plotted as a function of frequency for two antennas: antenna 11 (left) and 37 (right). The solutions are evaluated from three observations taken on three consecutive nights, at LST $= 22.7$ hours. The deviation of the solutions from the mean across the three days are displayed in the corner box for both antennas.}
\label{fig:calsoln}
\end{figure}

Next, we analyze the stability of the calibration solutions over multiple days. To do this, we compare the solutions at a common Local Sidereal Time (LST) across multiple consecutive days during the observation period. The analysis reveals deviations ranging from $4\%$ to $10\%$ from the mean solutions, reflecting their overall quality. Figure~\ref{fig:calsoln} shows the normalized amplitudes at an LST of 22.7 hours, with deviations of approximately $2-3\%$ observed in two individual antennas. Since the deviation is small and consistent across the three nights, it can potentially be attributed to time-dependent errors on the MWA primary beam \citep{Line2018} or changes in the ionospheric activity.


 The solutions are then applied to the pre-processed visibilities from Section~\ref{sec:flagging} to form calibrated visibilities.

\subsection{Foreground subtraction}
\label{sec:foreground_subtraction}
As demonstrated by \citet{Chege2022}, correcting for ionospheric turbulence in MWA data reduces contamination in the 21~cm power spectra. We correct the phase offsets for the 1,000 brightest sources, as they have enough signal-to-noise to yield accurate offset values \citep{Jordan2017}. Including fainter sources in the ionospheric correction could deteriorate the subtraction due to the large uncertainties associated with their offset calculations \citep{Chege2022}. The remaining 7,000 sources are subtracted using positions from the calibration source catalog described in Section~\ref{sec:calibration}. Foreground avoidance is then handled in the power spectrum estimation step.


\subsection{Data Quality Assurance}
\label{sec:data_quality_assurance}
In \citet{Nunhokee2024}, we introduced a statistical quality assurance framework designed to derive various metrics for identifying and mitigating systematics in MWA data observed with the Phase I configuration. These metrics are based on visibilities, calibration solutions, and images, enabling the detection of outlying observations and anomalies across time, frequency, and antennas. For this study, we have incorporated observations from the Phase II configuration and made several adjustments to adapt the framework accordingly.

Below are some of the primary data quality metrics used for this paper:
\begin{enumerate}
\item Data Quality Issues: The MWA Monitor \& Control system logs faults that occur during telescope operations.\footnote{This information is publicly accessible via the MWA Table Access Protocol (TAP) database at \url{https://mwatelescope.atlassian.net/wiki/spaces/MP/pages/24970532/MWA+ASVO+VO+Services}.} For this work, we focus on observations affected by corrupted data files, issues complicating preprocessing, or a high proportion of dead dipoles and bad tiles. Tiles with more than two dead dipoles and observations over $12\%$ bad tiles (equivalent to the number of tiles corresponding to two receivers)  are excluded from the analysis.


\item Flagging Occupancy: The occupancies are calculated relative to the total number of visibilities in the dataset and are classified into three types: Preprocessed Occupancy, AOFlagger RFI Occupancy, and SSINS Total Occupancy. Preprocessed Occupancy represents the proportion of data flagged during preprocessing, while AOFlagger RFI Occupancy denotes the fraction of visibilities identified as RFI by AOFlagger. SSINS Total Occupancy, derived using the SSINS algorithm, is further divided into three main categories: SSINS Narrowband RFI, SSINS Digital Television (DTV), and SSINS Streak (for more details see \citealt{Wilensky2019}). To maintain data quality, specific thresholds are applied in this work:

\begin{tabular}{c|c}
    Preprocessed occupancy & 50\%\\
    AOFlagger RFI occupancy & 3$\%$\\
    SSINS occupancy & 25\%\\
    SSINS narrowband RFI & 6$\%$\\
    SSINS DTV  & 5$\%$\\
    SSINS Streak & 3.1$\%$\\
\end{tabular}
\item $P_{win}(\textrm{unsub})$: Average power in $Jy^2$ evaluated using delay transformed visibilities from the wedge region. This metric quantifies the power present in the window prior to any foreground subtraction, where the cosmological 21~cm signal is expected to reside. Significant power in this region indicates leakage from foregrounds, which could introduce bias into the final results.

\item $\frac{P_{win}}{P_{wed}}(\textrm{unsub})$: Ratio of the power evaluated from delay transformed visibilities in the window region to the wedge.This ratio assesses the relative amount of foreground leakage into the window compared to the wedge. A high value suggests substantial contamination of the window by foregrounds, potentially caused by instrumental effects such as polarization leakage.

\item $P_{\small{wed}}(\frac{\textrm{sub}}{\textrm{unsub}})$: Ratio of the power confined into the wedge after foreground subtraction to before subtraction. This metric evaluates the effectiveness of the subtraction process by quantifying the fraction of foreground power removed from the wedge. Lower values indicate more successful subtraction.

\item $P_{\small{win}}(\frac{\textrm{sub}}{\textrm{unsub}})$: Ratio of the power confined into the window after foreground subtraction to before subtraction. Analogous to the metric for the wedge, this ratio reflects the effectiveness of foreground subtraction specifically within the window region, where minimal residual power is desired to preserve the integrity of the 21~cm signal.

\item $V_{\textrm{rms}} (\textrm{unsub})$ : Root mean square evaluated from the image pixels constructed using Stokes $V$ visibilities before foreground subtraction. This metric quantifies circular polarization in the observations. As no bright Stokes $V$ sources are expected in the field of view, the RMS should ideally be minimal. Elevated values may indicate instrumental leakage.

\item $\frac{S_{V}}{(S_{\small {EW}} + S_{\small{NS}})}$: Ratio of flux density $S$ of the brightest source in the field of view $PKS0023-026$ extracted from Stokes $V$ to sum of its flux densities extracted across East--West and North--South polarizations. This ratio evaluates instrumental leakage from Stokes $I$ into Stokes $V$. The denominator approximates the pseudo-Stokes $I$, and a significant non-zero ratio suggests polarization leakage, likely due to instrumental effects.

\item $(S_{\small{EW}} , S_{\small{NS}})$: Difference between flux densities $S$ across East--West and North--South polarizations. In an ideal scenario, flux densities should be nearly identical across both polarizations. Large discrepancies indicate the presence of instrumental systematics.

\item $S_{\small{EW}}(\frac{\textrm{sub}}{\textrm{unsub}})$: Ratio of $S$ extracted along East--West polarization after subtraction to before subtraction. This metric assesses the effectiveness of foreground subtraction by quantifying the residual flux from the brightest source in the East--West polarization. A value close to zero indicates successful subtraction.

\item $S_{\small{NS}}(\frac{\textrm{sub}}{\textrm{unsub}})$: Ratio of $S$ extracted along North--South polarization after subtraction to before subtraction. Similar to the East--West case, this metric quantifies the residual flux from the brightest source in the North--South polarization. Low values imply effective subtraction and minimal residuals.

\end{enumerate}

The cut-off threshold for determining which observations need to be discarded is estimated by using the 3$\sigma$ and interquartile rule on the distribution of the individual metrics across observations. The distributions are utilized as a guide on the rule to be applied. 


\subsection{Power Spectrum}
\label{power_spectrum}
The power spectrum $P(k)$ is defined as the Fourier transform of the two-point correlation function $\xi(\vec{r})$
\begin{equation}
P(|\vec{k}|) = \xi(\vec{r})e^{(-2\pi i\vec{k}.\vec{r})} d^3\vec{r},
\end{equation}
where $k$ is the wavenumber representing different spatial scales \citep{Trott2016_chips}. The angular scales $(u, v)$ in the interferometer plane are mapped onto perpendicular Fourier modes $k_{\perp}$ and the spectral modes are mapped onto line-of-sight Fourier modes $k_{||}$. The ability to relate interferometric measurements directly to Fourier modes makes the power spectrum a particularly powerful tool for analyzing 21~cm hydrogen line observations, as it provides a direct link between the observational data and the underlying cosmic structure \citep{Furnaletto2006, Furlanetto2016}. 

We construct the power spectra from the observations across three overlapping bands of 15.36~MHz bandwidth each: ($170.875$--$186.235$~MHz), ($177.275$--$192.635$~MHz), ($182.395$--$197.755$~MHz), corresponding to central redshifts of $z=7.0,\, 6.8,\,  6.5$, and for both polarizations. Note that the bottom 3.68~MHz of the full $30.72$~MHz band is not used due to discontinuities in the spectral response caused by missing data over a period in 2017.
Although the bands have significant overlap and wide individual bandwidths, the application of a Blackman-Harris spectral taper effectively reduces the bandwidth to approximately $9$~MHz, ensuring that the bands remain mostly independent.

The power spectra are computed using the standard Cosmological HI Power Spectrum (\textit {CHIPS}; \citealt{Trott2016_chips}) pipeline that was employed for the most recent large data integration \citep{Trott2020}, with an update to inpaint missing data from the MWA's coarse channelisation with low $k$ modes. In brief, visibilities with a common phase centre are gridded onto a $uv$-plane using a cell size of 0.5$\lambda$, with a 4-term Blackman-Harris gridding kernel that is matched to the beam size in Fourier space. No foreground statistical subtraction is performed (as had been used in \citealt{chips2016}); only the direct subtraction of a component of the compact sky model during the Hyperdrive calibration stage. The line-of-sight spectral analysis employed a weighted Fourier Transform. A non-uniform FFT was trialled, but the regularity of the missing spectral channels led to poor performance, and this was not used further. We also trialled an inpainting technique proposed by \citet{Chen2025} for HERA data that uses a Discrete Prolate Spheroidal Set (DPSS) basis set to inpaint missing data. The DPSS basis represents the solution to an eigenvalue problem for a filter in $k$-space that has a sharp cutoff at a specified delay. The inpainting technique requires additional work to tailor it to the particular features of the MWA bandpass, and we leave this exploration for future work.

We produce cylindrical (2D) power spectra as a final diagnostic for identifying any particular foreground leakage that has only manifested in the final large data integration. Spherically-averaged (1D) power spectra are generated for the final cosmological analysis\footnote{We used the following cosmological parameters: Hubble constant: $H0=100\,h$ km/s/Mpc, matter density parameter: $\Omega_{m}=0.272$, 
 baryon density parameter: $\Omega_b = 0.046$ and dark energy density parameter: $\Omega_{\Lambda} = 0.682$ \citep{Guo2013}.}, focusing on the least contaminated wavenumbers in the foreground avoidance window.
A minimum $k_\parallel$ value of $0.11\,h$ Mpc$^{-1}$ is used to avoid foreground-dominated modes, as well as a wedge cutoff of the horizon line, as defined in \citet{dillon15} and \citet{Barry2019a}. The noise includes both thermal and sample variance. Negative powers are reported as 2$\sigma$ noise. In this way, reducing the wavenumbers used for the final averaging will also increase the sample variance, and therefore these cuts are appropriately reflected in the uncertainty budget.


\begin{figure}
\includegraphics[width=1.0\linewidth]{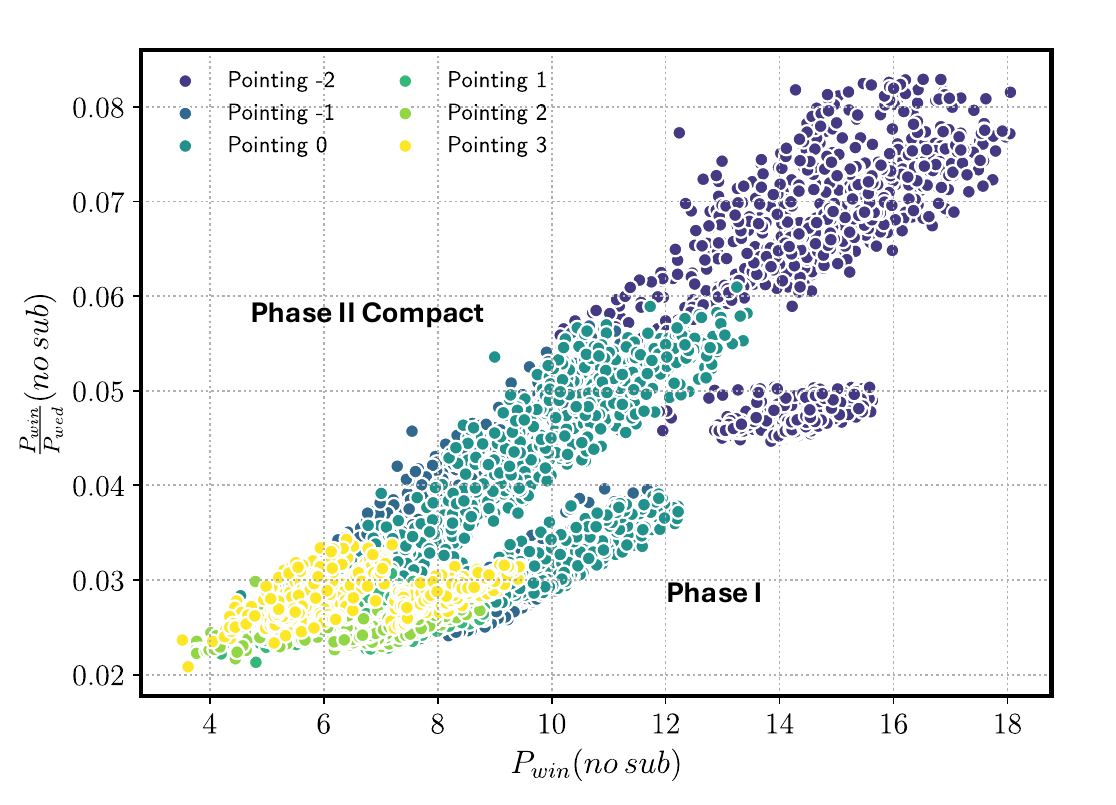}
\caption{Marginal distribution of two quality metrics $\frac{P_{win}}{P_{wed}}$ versus $P_{win}$, derived from delay-transformed visibilities, illustrating the systematic discrepancy between observations from Phase I and Phase II instrument configurations. Each point represents an observation, and colors correspond to the pointing directions listed in Table~\ref{tab:telescope_pointing}.}
\label{fig:phaseI_II_dist}
\end{figure}

\section{Results}
\label{sec:results}
This section explores the results obtained from the implementation of various statistical metrics and presents the power spectrum results from the final integration set.

\subsection{Data Filtering}
The derived quality metrics discussed in Section~\ref{sec:data_quality_assurance} reveal a clear discrepancy between the two array configurations. A marginal plot of two such metrics --$\frac{P_{win}}{P_{wed}}$ and $P_{win}$, evaluated from visibilities prior to foreground subtraction is shown in Figure~\ref{fig:phaseI_II_dist}. Each point represents an observation, with colors indicating different pointing directions. Observations clustered in the lower region of the plot correspond to Phase I configurations, while those with higher magnitudes originate from Phase II. This trend is consistently observed across all metrics, highlighting a distinct difference in systematics between the two configurations. Consequently, we treated the metric distributions separately for thresholding purposes. 
Estimating thresholds based on the combined distribution would result in both false positives and false negatives. As a result, $48\%$ of the observations were discarded, leaving us with 314.4~hours (9,433 observations) of usable data.

\begin{figure}
\includegraphics[width=1.0\linewidth]{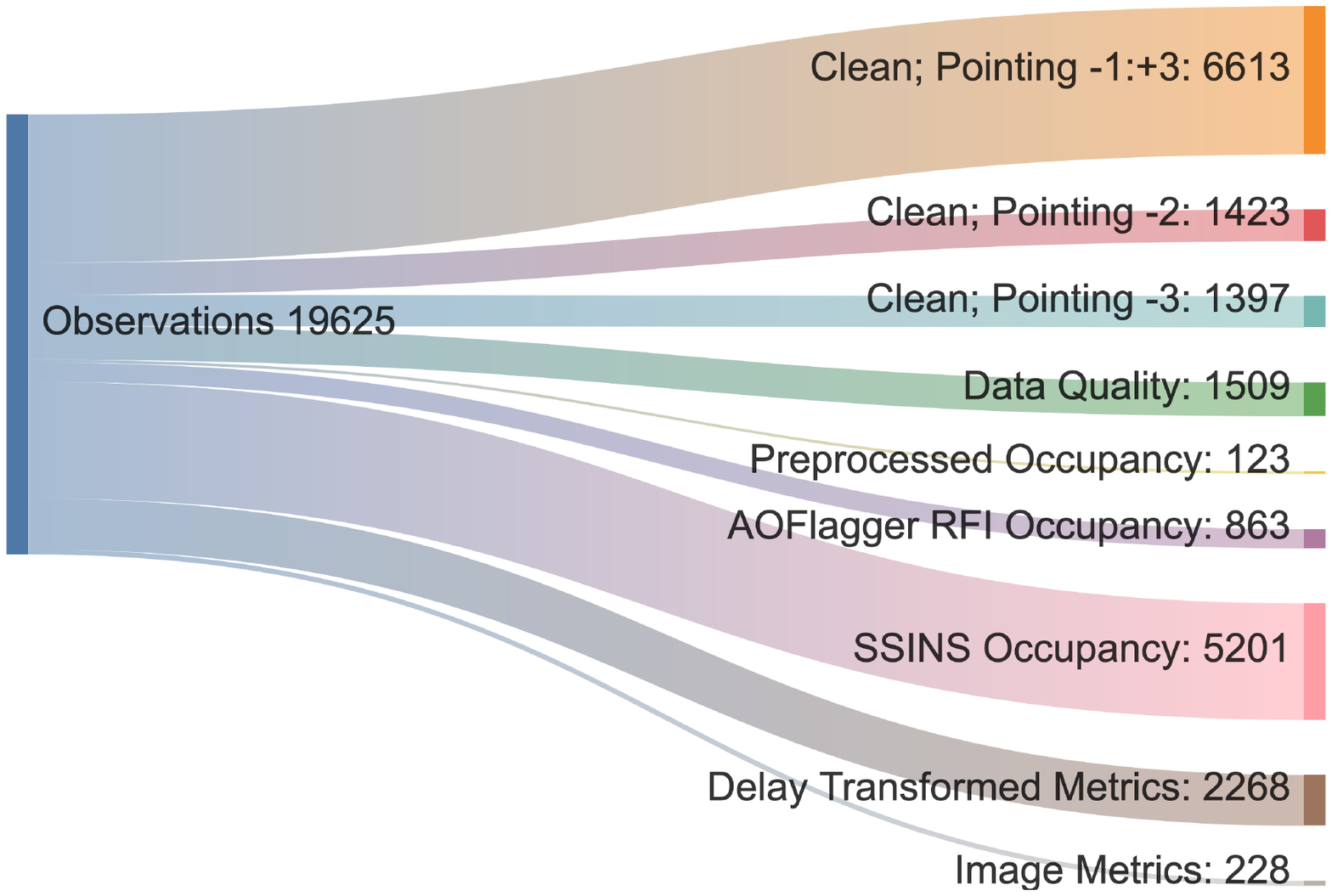}
\caption{A Sankey diagram illustrating the progression of discarded observations at each intermediate stage, as shown in Figure~\ref{fig:eor_pipeline}. The Delay-Transformed Metrics and Image Metrics correspond to metrics 3--5 and 6--10, respectively, as detailed in Section~\ref{sec:data_quality_assurance}.}
\label{fig:sankey_breakdown}
\end{figure}

Figure~\ref{fig:sankey_breakdown} illustrates the breakdown of discarded observations according to the applied metrics, with the highest rejection rate of approximately $27\%$ attributed to SSINS filters. This significant fraction is likely due to emissions from satellites or digital television (DTV) signals detected by the instrument, particularly those obscured beneath the per-baseline thermal noise, as discussed in \citet{Wilensky2019, Wilensky2023}.

\begin{figure}
\centering
\includegraphics[width=1.0\linewidth]{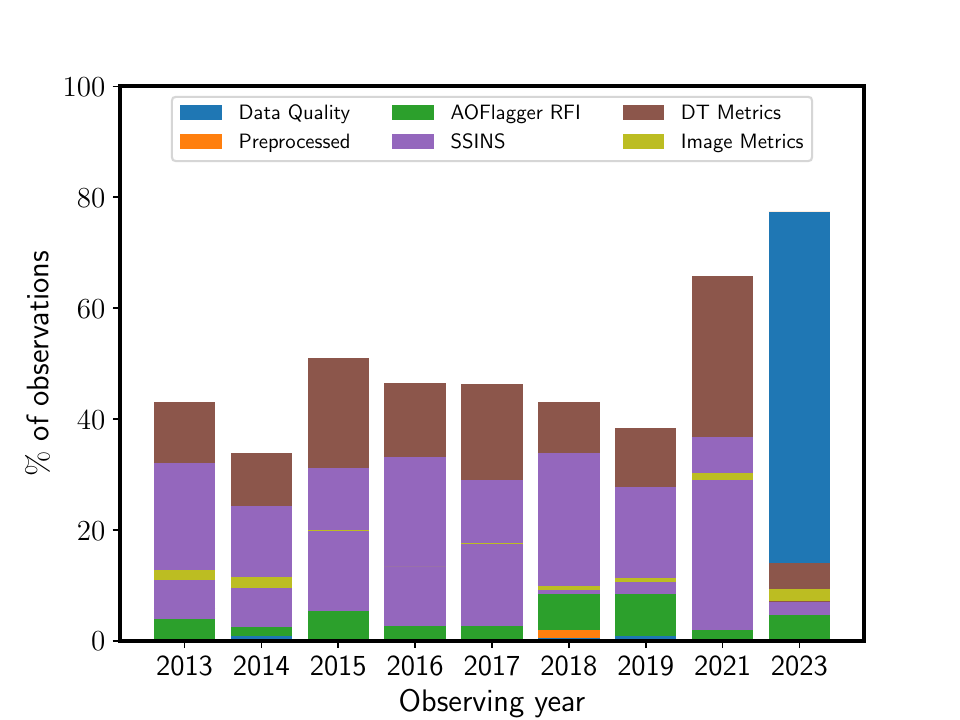}
\caption{Percentage of observations discarded due to quality issues across different years. Observations prior to 2016 were conducted using the Phase I configuration, while later years used the Phase II compact configuration. The notably low number of clean datasets in 2023 is due to widespread tile malfunctions and a limited number of observations.}
\label{fig:metrics_breakdown}
\end{figure}

The yearly breakdown of discarded observations for individual metrics is presented in Figure~\ref{fig:metrics_breakdown}. The bar chart highlights the dominant effect of RFI across all years. Observations conducted up to 2015 used the Phase I configuration, while those from 2016 onward employed the Phase II compact configuration. The percentage of discarded observations across the various metrics remains relatively consistent between the two configurations. In 2023, however, a significant number of tiles were found to be malfunctioning during the preprocessing stage, resulting in only 17 clean datasets. This deviation is also attributed to the limited number of observations available for that year.

Clearly, observations from the Phase II compact configuration were more dominated by systematics, with 55\% of observations rejected, compared to 47\% in Phase I. This is further supported by the direction-independent calibration convergence values. The mean convergence values in Phase II are not only higher but also exhibit a greater dynamic range on the order of 3 compared to Phase I, which shows a scale on the order of 2. This suggests that the calibration process struggled more with the Phase II compact observations. We also independently generated power spectra from a subset of 400 non-rejected observations for both configurations. Despite the thermal noise in the Phase II compact configuration being lower by a factor of 0.4, consistent with its improved sensitivity, the resulting power spectra show negligible differences.

 
\subsection{21~cm Upper Limits}
\label{sec:upper_limits}
\begin{figure*}
\includegraphics[width=1.0\linewidth]{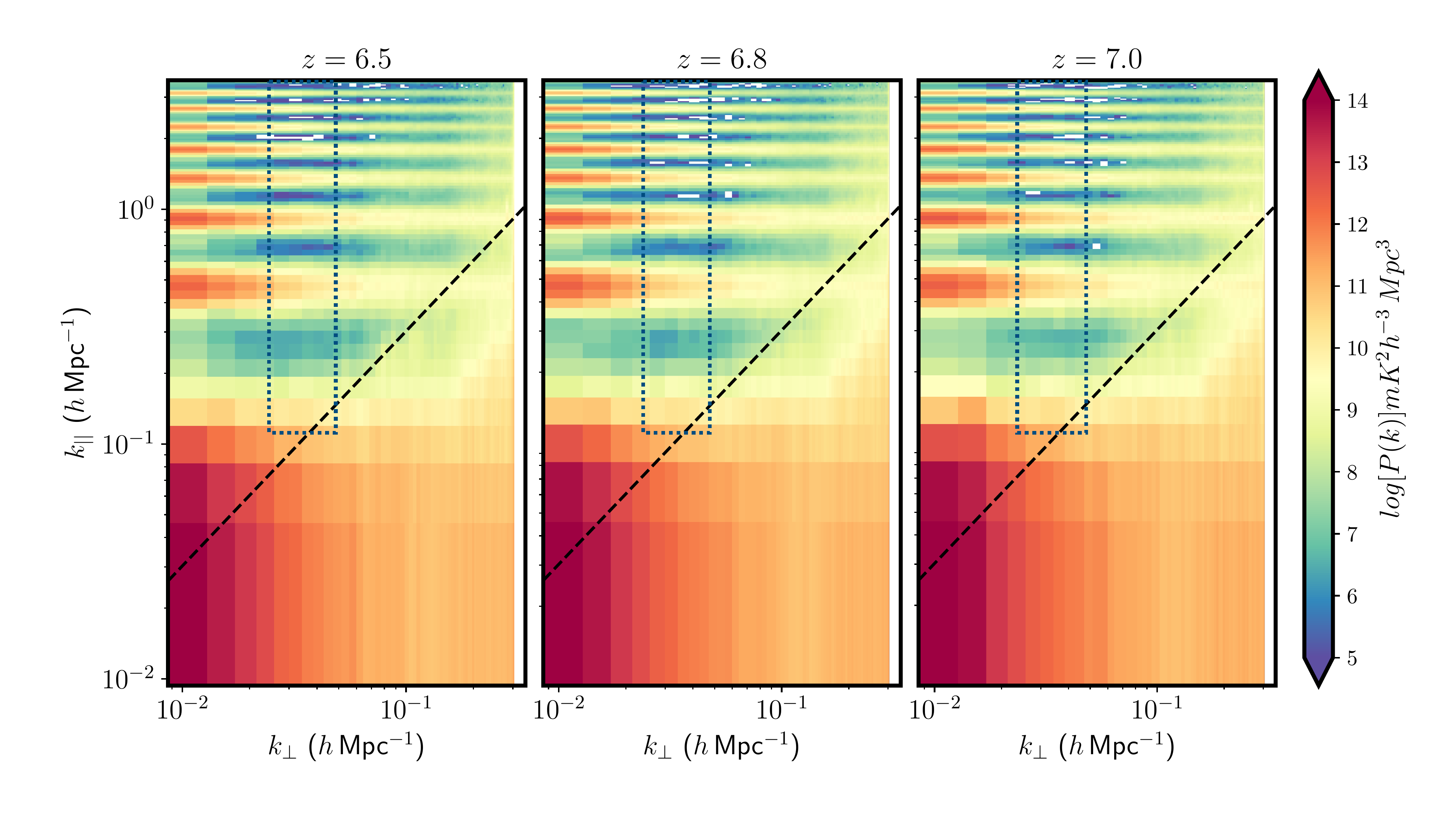}
\caption{Cylindrical (two-dimensional) power spectra generated from 8,036 observations at $z=6.5$, $6.8$ and $7.0$ and along the East--West polarization. The dashed black lines represent the horizon limit. The dashed blue box indicates the region enclosed by $0.025< k_{\perp} < 0.045 \, h$ Mpc$^{-1}$ and $k_{||} > 0.11 h$ Mpc$^{-1}$.}
\label{fig:power2d}
\end{figure*}

In addition to the aforementioned filtering, observations from pointings at -3 are excluded due to the strong presence of the Galaxy near the horizon. The aliasing effects of the Galactic plane, as discussed in \citealt{Barry2024}, are particularly pronounced for this pointing, as evidenced by the significant disparities in the metrics outlined in Section~\ref{sec:data_quality_assurance}. Observations from pointings -2 to 3, comprising 8,036 individual measurements (Phase I: 4,050 \& Phase II compact: 3,986) and totaling 268 hours (Phase I: 135~hours \& Phase II compact:  132.9~hours), were therefore combined to form the final power spectra. The data, phased to a common phase center, were coherently averaged using an optimal weighting scheme within a consistent framework. Signal decoherence due to variations in the beam response across pointing centers was found to be negligible, as noted by \citet{Trott2020}. For illustration, Figure~\ref{fig:power2d} presents the cylindrical power spectra from these observations for the East–West polarization. The residual foregrounds are predominantly confined to the lower portion of the power spectrum, commonly referred to as the wedge. The region above the black horizon line denotes the EoR window, where the 21~cm cosmological signal is expected to reside.


\begin{figure*}[h]
\centering
\includegraphics[width=1.1\linewidth]{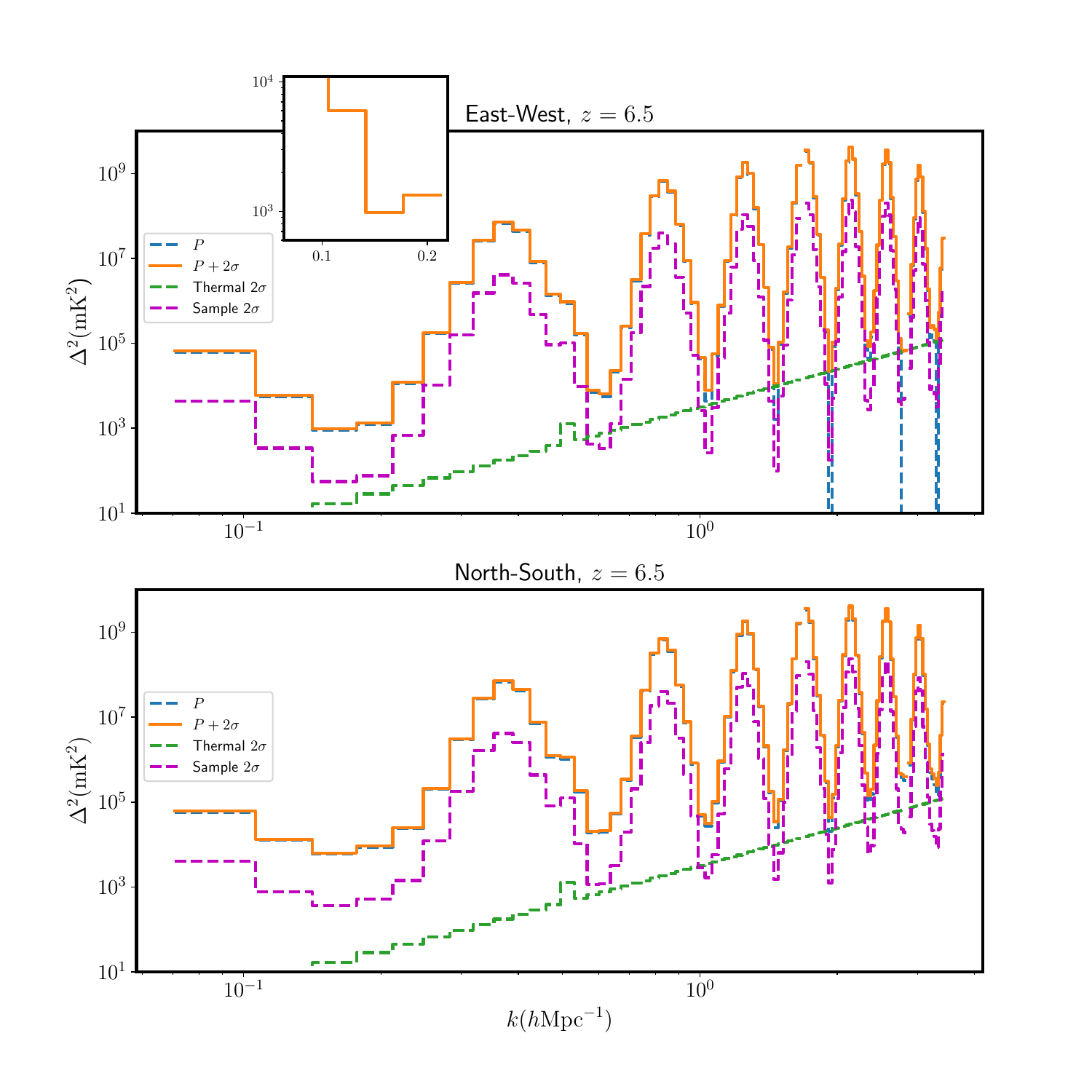}
\caption{Spherically-averaged 2$\sigma$ upper limit power spectra at $z=6.5$ across the East--West (top) and North--South (bottom) polarizations in orange from 6635 observations and $k_{\perp}: 0.020-0.045\, h\,$Mpc$^{-1}$. The North-South oriented dipoles are more susceptible to contamination from the Galaxy near the horizon \citep{Barry2024}, leading to poorer results. The inset on the top panel highlights the $k-$mode where our best upper limit resides. Also shown are the thermal noise power (green) and sample variance (magenta) that together comprise the associated error. }
\label{fig:z6pt5}
\end{figure*}

We then generate the spherically-averaged 2$\sigma$ upper limit power spectra at $z=6.5, 6.8$ and $7.0$. The averaging is performed over the range $0.025\,h$ Mpc$^{-1}$ $< k_{\perp} < 0.045\, h$ Mpc$^{-1}$ and $k_{\parallel} > 0.11\,h$ Mpc$^{-1}$, as indicated by the blue enclosed region in Figure~\ref{fig:power2d}. This selection aims to minimize contamination from the foregrounds and systematics. Lower $k_{\perp}$ modes are excluded due to their dominance by systematics, as seen in Figure~\ref{fig:power2d}. Following \citet{Li2019}, the upper bound is determined based on the $uv$ sampling of our data. This analysis incorporates observations from both Phase I and Phase II configurations, ensuring comprehensive $uv$ coverage. 
Figure~\ref{fig:z6pt5} displays the spherically averaged 2$\sigma$ upper limit power spectra at $z=6.5$ for the East--West (top) and North--South (bottom) polarizations (orange, solid lines). The thermal noise power is shown in green, while the sample variance is depicted in magenta; together, these represent the total error. Table~\ref{table:results} presents the upper limits at $z=6.5$, $6.8$, and $7.0$ for both polarizations, along with the sample variance and thermal noise. For a complete list of power spectrum values, refer to Table~\ref{tab:allz_results}.

\subsection{Statistical Tests on the 21~cm upper limits}
\begin{figure}
\includegraphics[width=1.0\linewidth]{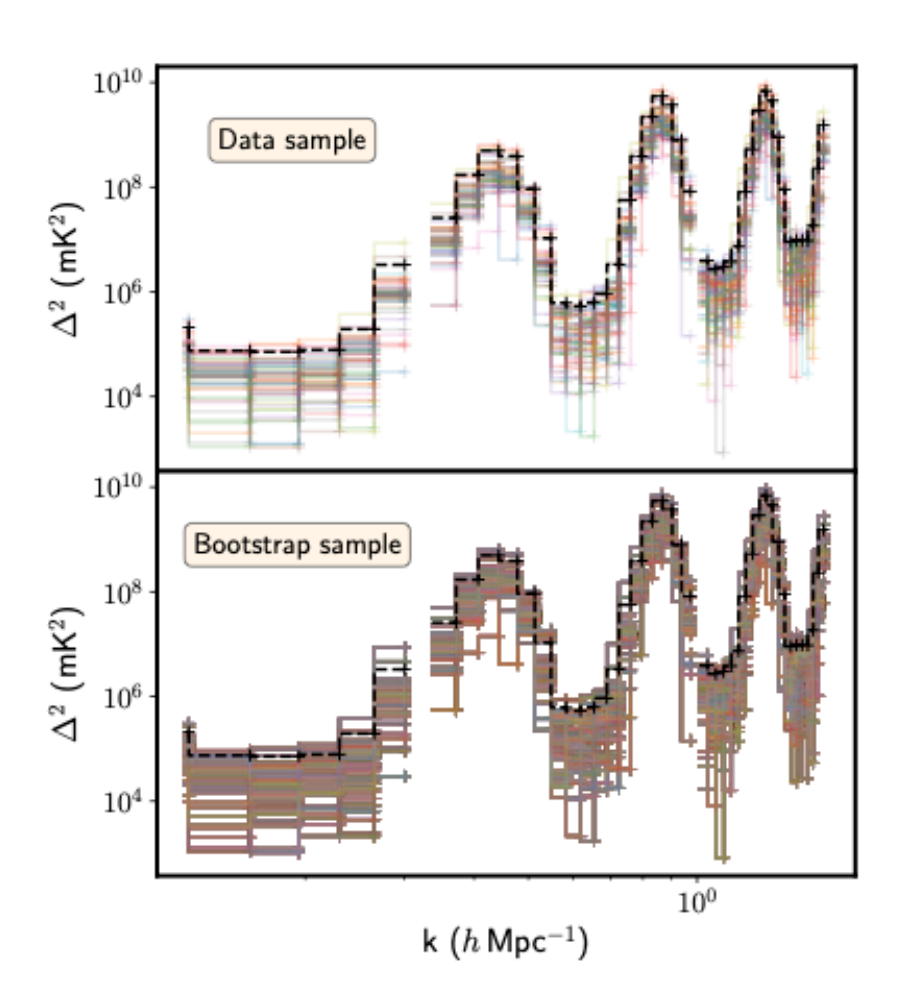}
\caption{{\textit{Top}}: Spherically-averaged power spectra from subsets of $80$ observations at $z=6.5$. {\textit{Bottom}}: Bootstrapping on the residuals obtained by subtracting the realized samples from the mean estimate from a sample of 6,000.}
\label{fig:bootstrap}
\end{figure}

We perform statistical tests to establish confidence in our upper limits. As a first test, we evaluate the spherically-averaged power spectra using the same cuts defined in Section~\ref{sec:upper_limits} on 100 subsets of 80 observations, displayed in the top panel of Figure~\ref{fig:bootstrap}. The choice of the number of observations in the subset is based on the computing limitation of our pipeline. The $2\sigma$ error on the mean distribution, derived from the empirical distribution of the spherically averaged power spectra within the subsets, is represented by the black dotted line. Most of the spherically-averaged power spectra lie within the error boundaries, with no indication of any bias.

We also perform bootstrapping on the noise from the one-dimensional spherically-averaged power spectra to assess whether any observations remain dominated by systematics. To calculate the noise for individual subsets, we subtract the power of each subset from the mean power across all 80 observation subsets.

For bootstrapping, we draw samples from the spherically-averaged power spectra, with each value returned to the pool of available values before the next draw. This allows the same data point to be selected multiple times within a single resampling. 
The $2\sigma$ uncertainty over the mean, derived from $6,000$ bootstrapped samples at redshift $z=6.5$, is shown in the lower panel of Figure~\ref{fig:bootstrap}. We found no evidence of anomalies in the noise distributions.

\begin{table*}
\centering
\begin{tabular}{|c||c|c|c|c|c|c|c|c|c|}
\hline 
$z$ & $N_{hr}$ & $N_{obs}$ & $k$ (h/Mpc) & $k_\perp$ cuts (h/Mpc) & P$_{UL}$ (mK$^2$) & P$_{therm}$ (mK$^2$) & P$_{sample}$ (mK$^2$) & Pol. \\ 
\hline \hline
6.5 & 268 & 8036 & 0.18 & 0.025--0.045 &  (30.2)$^2$ & (3.7)$^2$ & (7.4)$^2$ & EW\\
6.8 & 268 & 8036 & 0.18 & 0.025--0.045 &  (31.3)$^2$ & (4.1)$^2$ & (7.5)$^2$ & EW \\
7.0 & 268 & 8036 & 0.21 & 0.025--0.045 &  (39.1)$^2$ & (6.1)$^2$ & (9.6)$^2$ & EW \\
\hline
6.5 & 268 & 8036 & 0.21 & 0.025--0.045 &  (39.2)$^2$ & (4.8)$^2$ & (9.5)$^2$ & NS \\
6.8 & 268 & 8036 & 0.18 & 0.025--0.045 &  (79.3)$^2$ & (4.1)$^2$ & (18.9)$^2$ & NS \\
7.0 & 268 & 8036 & 0.18 & 0.025--0.045 &  (55.3)$^2$ & (4.6)$^2$ & (13.6)$^2$ & NS \\

\hline 
\end{tabular}
\caption{Best 2$\sigma$ upper limits for each redshift and each polarization (P$_{UL}$), where both the thermal noise and sample variance are included in the error budget (values reported are 2$\sigma$ for each error term). 
Columns are redshift, number of hours, number of 2-min snapshot observations, $k$ mode of the lowest power, $k_\bot$ cuts used, 2$\sigma$ power upper limit, thermal noise power, sample variance, polarization.}\label{table:results}
\end{table*} 



\section{Validation}
\label{sec:validation}
Validation of the reported results is crucial to demonstrate their robustness. We take two approaches to the validation, both of which support different components of the work.

Firstly, we employ a full end-to-end simulation of the data, instrument and analysis pipeline to show that the pipeline produces 21~cm power levels that are input from the beginning, in the presence of noise, diffuse emission and compact foregrounds. This validation pipeline and its results are reported in \citet{Line2024}, and define an overall data normalization correction that we need to include to restore full cosmological power \citep{Liu2014, Barry2019a}. This validation work uses a realistic 21~cm signal model \citep{Greig2022}, and matches the observational strategy used by the MWA for the data used in this work. The validation calculation is performed again with the final pipeline used in this work, to ensure that the recovered 21~cm power matched that in the test dataset.



The second approach employs a comparison between an independent analysis pipeline and our pipeline on a subset of the data. This approach has been previously used by the international MWA EoR Collaboration to validate results \citep{jacobs16,beardsley_first_2016,Barry2019b,li_first_2019}. Here, we use the Fast Holographic Deconvolution (FHD; \citealt{sullivan_fast_2012}) and $\epsilon$ppsilon pipeline \citep{Barry2019a} for independent analysis. 

The FHD/$\epsilon$ppsilon pipeline, whilst similar to the Hyperdrive/CHIPS pipeline, uses a variety of different techniques in analysis. 
\begin{itemize}
    \item Calibration in FHD/$\epsilon$ppsilon uses auto-correlations to constrain a normalised spectral amplitude, and all catalog sources down to the horizon to constrain the phase and absolute amplitude. The spectral variation of the calibration solutions are then averaged over each season and pointing (see \citealt{Barry2019a}).
    \item Calibrated visibilities are gridded with a Gaussian-decomposition of the beam to avoid aliasing \citep{barry_role_2022} -- for simplicity, only the Gaussian recreating the FWHM of the FEE pattern was used for this validation.
    \item Images are made at 160~kHz resolution in HEALPix projection \citep{gorski_healpix:_2005} extending nearly to the horizon to avoid aliasing. These images are then coherently integrated in this basis.
    \item The coherent image is transformed back to the Fourier basis. Maximum-likelihood estimates for the mean, observed noise, and uncertainties are propagated from the initial variance of the visibilities assuming fully independent measurements. These are then compared to an expected noise calculated from a fully independent noise simulation and propagated through the pipeline. The comparison yields near-unity, suggesting a lack of significant cross-correlations in measurements and robustness in the calculated uncertainties. 
\end{itemize}  

A subset of $335$ observations, chosen to be representative of the full sample, is processed through both pipelines. To keep the comparison solely between the analysis methodologies, some aspects are constrained to be the same. This includes RFI flagging, individual tile flagging, input sky catalogues, binning schemes, and Fourier transform window functions. The thermal noise from the data subset for each analysis is thus constrained to be comparable.

\begin{figure*}
\includegraphics[width=1.0\linewidth]{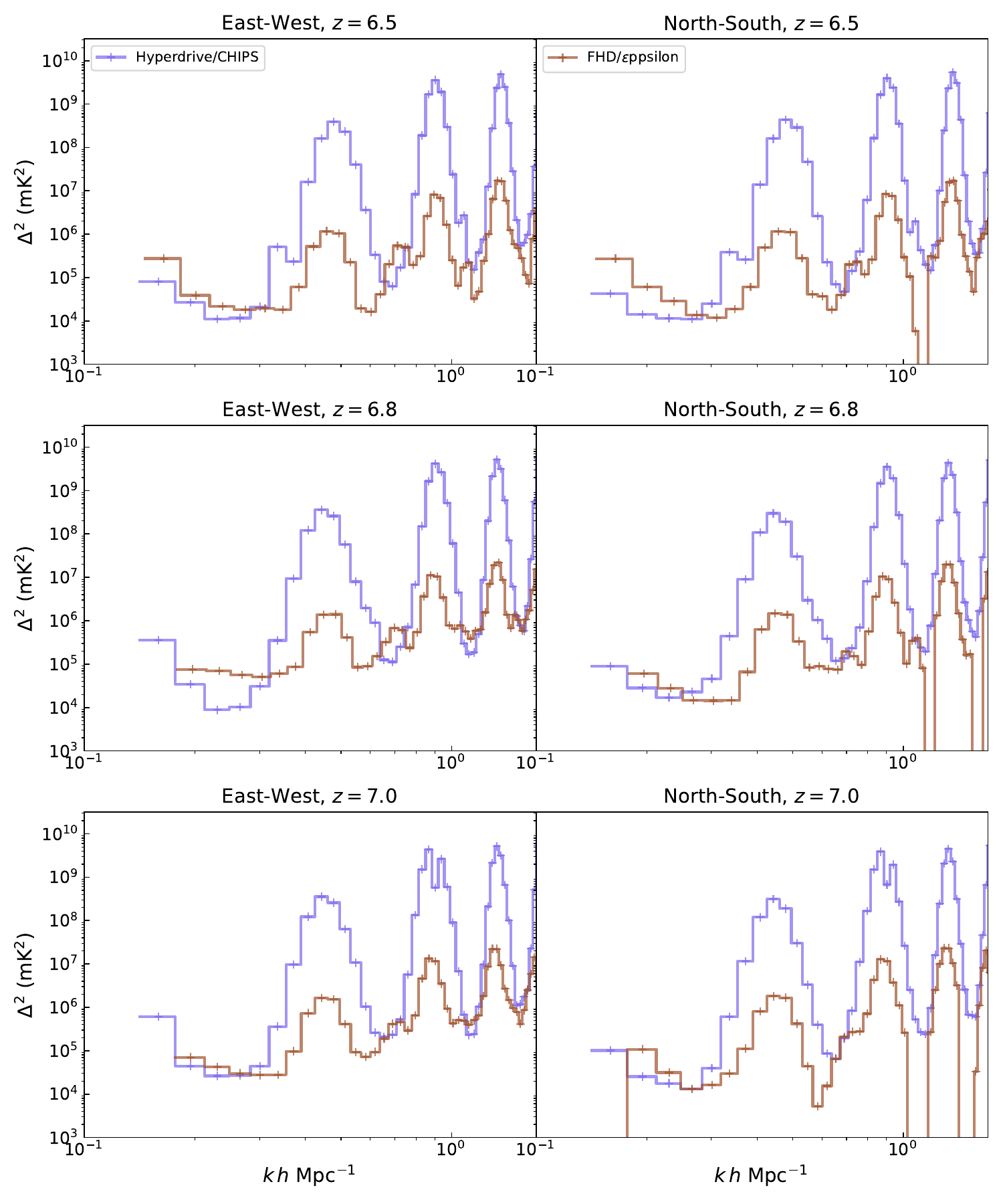}
\caption{Comparison of the one-dimensional EoR power spectra between Hyperdrive/CHIPS and FHD/$\epsilon$ppsilon for East--West and North-South polarizations at redshifts $z=6.5$, $z=6.8$ and $z=7.0$. There is general agreement across redshift and wavenumber for modes that are not affected by aliasing of foreground contamination due to the MWA's bandpass.}
\label{fig:validation}
\end{figure*}

Figure~\ref{fig:validation} shows the measured power for redshifts z = 6.5, z = 6.8, and z = 7.0 and for each polarization from each pipeline. We set  $0.015 < k_{\perp}< 0.06$\,$h$Mpc$^{-1}$, and $k_{||}>0.11\,h$ Mpc$^{-1}$, and $k_{||} > 2 \times k_{\perp} + 0.08\,h$Mpc$^{-1}$. This avoids the majority of foreground-contaminated modes and poorly sampled $uv$-modes while retaining as much data as possible.  

The main difference between the validation results from Hyperdrive/CHIPS and FHD/$\epsilon$ppsilon is the strength of the coarse band flagging contamination (approximately $k$ = 0.043, 0.09, and 1.3 $h$Mpc$^{-1}$). The polyphase filter bank of the MWA requires flagging 40~kHz at the beginning and end of each 1.28~MHz band. Hyperdrive/CHIPS averages to 80\,kHz frequency resolution throughout the post-calibration analysis. FHD/$\epsilon$ppsilon averages up to 160~kHz when creating images. Aliasing from the polyphase filter bank flagging depends on the final frequency resolution, and thus aliasing contamination is different between the two analyses. Other differences include contamination from the 150~m and 230~m cable reflection \citep{beardsley_first_2016} being more prevalent in FHD/$\epsilon$ppsilon, which may reflect that the per-field averaging is accidentally propagating cable residuals. 

Hyperdrive/CHIPS has significantly lower power in the East--West polarization of z = 6.8 at sensitive $k$-modes. We investigate this discrepancy by conducting several tests. FHD constructed power spectra in sequential groups of 50 observations and compared power levels generated from the best and worst sets, yielding comparable results in both Hyperdrive/CHIPS and FHD/$\epsilon$ppsilon.  This result reduces the likelihood that the observed discrepancy stems from the specific $50$ faulty observations that may be treated differently by the pipelines. We then assess whether calibration influences this behavior by calibrating a subset of the observations using Hyperdrive and processing them through the power spectrum pipeline in both FHD/$\epsilon$ppsilon and CHIPS. The resulting power levels remain consistent, ruling out calibration errors as the source of the discrepancy.

Notably, this behavior is only prominent at $z=6.8$, corresponding to a center frequency of 184.5~MHz, which aligns with the broadband digital TV Channel 7 (DTV7). This suggests that low-level residual radio frequency interference (RFI), undetected by SSINS, may be responsible. To investigate further, we process the data using EAVILS, a complementary flagging algorithm currently in development. EAVILS identifies contamination within the DTV7 band, confirming the presence of residual RFI. Building on this RFI contamination analysis, we conduct simulations using WODEN \citep{Line2022} to evaluate how both pipelines manage RFI contamination and to verify that CHIPS does not introduce signal decoherence when integrating datasets. Specifically, we simulate the following scenarios:
\begin{enumerate}
    \item We simulate five point sources within the main lobe, with flux densities ranging from 0.5~Jy to several Janskys, across 69 observations spanning two hours.

    \item We introduce a 7~Jy DTV7-like RFI, at the southern horizon during 17 randomly distributed observations throughout the night. If no mitigation is in place, this should lead to power spectrum contamination similar to that simulated in \citet{Wilensky2020}.
\end{enumerate}
We then generate power spectra for three cases: (i) the 69 RFI-free observations, (ii) the 17 RFI-contaminated observations, and (iii) the full dataset containing both RFI-free and RFI-contaminated observations. Our results show that when processing RFI-contaminated visibilities, FHD/$\epsilon$ppsilon produces power levels approximately an order of magnitude higher than CHIPS, whereas power levels remain consistent between the two pipelines for RFI-free observations. This discrepancy persists even after incorporating RFI-free observations, indicating that CHIPS more effectively suppresses horizon contamination without decohering the signal. The primary difference in power levels arises from the gridding kernel, which influences horizon power suppression in each pipeline. CHIPS uses a 2D Blackman-Harris taper as a gridding kernel; the kernel size is matched to the instrumental Fourier beam, with extensive testing undertaken to ensure no signal loss in the main lobe and correct power recovery. The taper's response in image space is to retain power within the main lobe, but heavily suppress power outside. This is in contrast to the single Gaussian reconstructing the FWHM of the main lobe used in this implementation of FHD/$\epsilon$ppsilon (recent published limits from FHD/$\epsilon$ppsilon  have used a Blackman Harris to suppress horizon emission, e.g., \citealt{Barry2019b, Wilensky2023}).
While this difference is minor, it is enough to indicate that the discrepancy in the data could be caused by horizon RFI suppression from differing gridding kernels.

Other polarizations and redshifts have equivalent measured power levels at sensitive $k$-modes, while FHD generally performs better at high $k$. The overall agreement indicates that there are no outstanding biases present in the Hyperdrive/CHIPS pipeline that are not present in the FHD/$\epsilon$ppsilon pipeline. Given the significant analysis differences, including calibration, gridding kernels, integration space, and noise calculation methods, this is an indication of robustness.

Some of the processing for Hyperdrive/CHIPS was carried out in collaboration with industry partner Downunder Geosolutions, as their resources were better suited for the task.

\section{Astrophysical interpretation}
\label{sec:inference}
To understand the implications of the 21-cm power spectrum upper limits from the MWA, we use a Bayesian framework based on {\it 21cmFAST} \citep{Mesinger2011MNRAS.411..955M,Greig2018MNRAS.477.3217G,Park2019MNRAS.484..933P,Murray2020JOSS....5.2582M}
to infer the properties of the high-redshift IGM and the galaxies that drive the 21-cm signal. {\it 21cmFAST} forward-models the 21-cm power spectra based on bulk galaxy properties parameterized as functions of the host halo mass. These include the stellar-to-halo mass ratio, UV ionizing escape fraction, star formation timescales, duty cycles, and X-ray spectral energy distribution (SED), controlled by 9 free parameters. Among these parameters, the soft-band X-ray luminosity per unit star formation rate (SFR) is most constrained by current 21-cm limits \citep{Greig2021MNRAS.500.5322G,Greig2021MNRAS.501....1G,HERACollaboration2022A,HERA2023}, as it determines the strength of IGM heating during the EoR. For computational efficiency, we employ the {\it 21cmFAST} emulator, {\it 21cmEMU} \citep{Breitman2024MNRAS.527.9833B}, that has sub-percent median accuracy 
among the prior range in all forward-modeled observables explored in this work. For additional details on {\it 21cmFAST} or {\it 21cmEMU}, see the referenced works.

We perform Bayesian inference using the {\it Dynesty} sampler \citep{Speagle2020MNRAS.493.3132S} under two scenarios: one that incorporates the MWA upper limits into the likelihood as an additional error functional form proposed by \citet{HERACollaboration2022A} to penalize models with power higher than the observed limits, and one that does not. We consider the East--West and North--South polarizations as independent constraints. Both analyses account for additional multi-wavelength EoR probes, including:
\begin{enumerate}
    \item The CMB optical depth measured by {\it Planck} \citep{Planck2020}, using a double-sided Gaussian likelihood \citep{Qin2020MNRAS.499..550Q};
    \item The galaxy UV luminosity function established by {\it HST} \citep{Bouwens2015ApJ...803...34B,Bouwens2016ApJ...830...67B,Oesch2018ApJ...855..105O}, following the likelihood formalism in \citet{Park2019MNRAS.484..933P}; and
    \item The neutral fraction ($x_{\rm HI}$) inferred from Lyman-$\alpha$ forests observed by {\it VLT} \citep{Bosman2022,DOdorico2023MNRAS.523.1399D} at $z=6$, 7, and 8, with values of $x_{\rm HI}$ taken from Table 2 of \citet{Qin2024arXiv241200799Q} and incorporated as Gaussian likelihoods.
\end{enumerate}
These complementary probes place robust constraints on the star formation efficiency and the UV ionizing escape fraction of high-redshift galaxies, reinforcing that reionization was primarily driven by faint, low-mass galaxies. However, these multi-wavelength observations are largely insensitive to the heating properties of the IGM, highlighting the unique role of 21-cm cosmology in probing IGM thermal evolution during reionization. 

\begin{figure*}[h]
\centering
\includegraphics[width=1\textwidth]{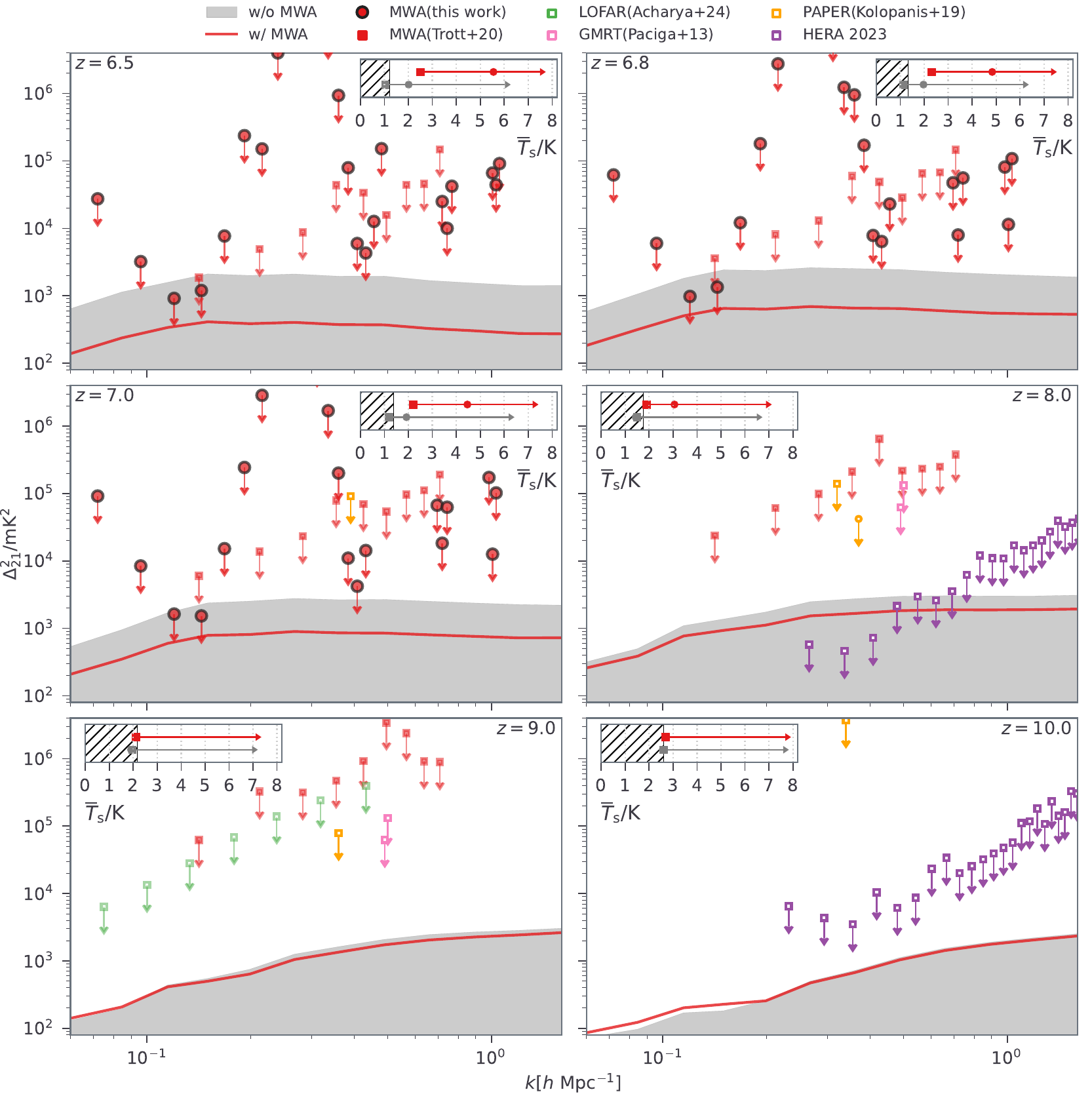}
\caption{The inferred power spectra with and without the MWA upper limits. The shaded region and solid curve indicate the upper bound of the 95\% C.I. of the posterior probability distributions. For clarity, only the deepest 2$\sigma$ limit between the East--West and North--South polarizations is shown at each wave-number. For comparison, we also include previous measurements from earlier MWA work \citep{Trott2020} as well as results from other experiments, including HERA \citep{HERA2023}, LOFAR \citep{Acharya2024MNRAS.534L..30A}, GMRT \citep{Paciga2013MNRAS.433..639P} and PAPER \citep{Kolopanis2019ApJ...883..133K} are also shown for comparison. In the corner of each subpanel, we present the 68\%  (circles) and 95\% (squares) C.Is. of the inferred IGM spin temperature where the hatched area indicates the adiabatic cooling limit. By excluding models with excessive power, MWA obtains the first evidence of a heated IGM at $z\sim6.5$--7. }
\label{fig:inference_ps}
\end{figure*}

Fig. \ref{fig:inference_ps} shows the inferred 21-cm power spectra at $z=6.5$ to 10, alongside the MWA results (this work and \citealt{Trott2020}) and other experiments including HERA \citep{HERA2023}, LOFAR \citep{Acharya2024MNRAS.534L..30A}, GMRT \citep{Paciga2013MNRAS.433..639P} and PAPER \citep{Kolopanis2019ApJ...883..133K} for comparison. We see a range of models with excessive power at $z=6.5$ to 7.0 are now ruled out by the MWA limits. These models, often referred to as ``cold reionization'', predict high 21-cm power spectra due to strong spatial fluctuations in the IGM ionized fraction and density which are amplified by minimal IGM heating. The exclusion of these models implies a more thermally evolved IGM than simple adiabatic cooling, with sufficient X-ray heating to suppress power at the observed redshifts. Extrapolating our results to higher redshifts where MWA constraints are weaker \citep{Trott2020}, we see the predicted power remains lower than (i.e. consistent with) most existing measurements\footnote{This includes the recent update from LOFAR where they further reduced the systemic noises using a variational autoencoder (see more in \citealt{Acharya2024MNRAS.534L..30A}).}, except for HERA\footnote{In other words, HERA still holds the deepest limit among all existing interferometers and, based on galaxy-evolutionary models such as {\it 21cmFAST}, provides the strongest constraints on the properties of high-redshift IGM and galaxies that drive the 21-cm signal\citep{HERA2023}} at $z=8$. 

In the inset of each panel in Fig. \ref{fig:inference_ps}, we also present the lower limits of the predicted gas spin temperature ($T_{\rm S}$). Without the MWA constraints, our parameter space is dominated by models with little or no heating, where the gas temperature reaches as low as the adiabatic cooling limit. However, incorporating the MWA limits eliminates these low-temperature models, yielding the first evidence for a heated IGM at $z=6.5$ to 7.0. The highest posterior density (HPD) 95\% confidence intervals (C.Is.) for the spin temperature are:
\begin{enumerate}
    \item $T_{\rm S}=2.5$ -- 797.8~K at $z=6.5$;
    \item $T_{\rm S}=2.3$ -- 758.6~K at $z=6.8$; and
    \item $T_{\rm S}=2.2$ -- 727.3~K at $z=7.0$.
\end{enumerate}

\begin{figure}[h]
\centering
\includegraphics[width=1\columnwidth]{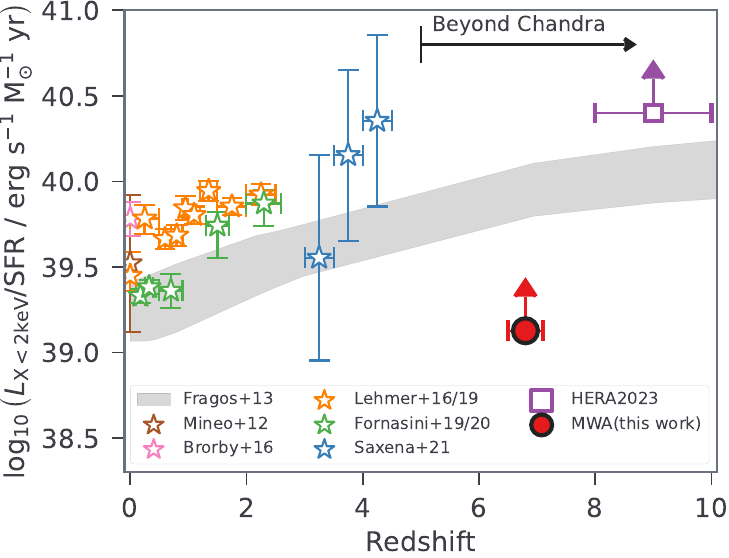}
\caption{The inferred soft-band X-ray luminosity per unit SFR from the MWA upper limits (68\% C.I.). Results using {\it Chandra} targeting low-redshift star forming galaxies \citep{Mineo2012MNRAS.419.2095M,Brorby2016MNRAS.457.4081B,Lehmer2016ApJ...825....7L,Lehmer2019ApJS..243....3L,Fornasini2019ApJ...885...65F,Fornasini2020MNRAS.495..771F,Saxena2021MNRAS.505.4798S} and HERA \citep{HERA2023} are shown for comparison. By excluding models with excessive power, MWA sets a lower limit of $L_{\rm X<2keV}/{\rm SFR}>10^{39}{\rm erg\ s^{-1}\ M_{\odot}^{-1} yr}$ at $z\gtrsim6.5$--7. }
\label{fig:inference_lx}
\end{figure}

Ruling out ``cold reionization'' models also has implications for the sources of IGM heating, which in {\it 21cmFAST} are assumed to be high mass X-ray binaries (HMXBs) in star-forming galaxies whose soft-band X-ray luminosity per SFR ($L_{\rm X<2keV}/{\rm SFR}$) governs heating in the early universe. In Fig. \ref{fig:inference_lx}, we present our updated constraints on $L_{\rm X<2keV}/{\rm SFR}$ derived from the MWA limits, together with previous results obtained from deep {\it Chandra} observations targeting low-redshift star forming galaxies \citep{Mineo2012MNRAS.419.2095M,Brorby2016MNRAS.457.4081B,Lehmer2016ApJ...825....7L,Lehmer2019ApJS..243....3L,Fornasini2019ApJ...885...65F,Fornasini2020MNRAS.495..771F,Saxena2021MNRAS.505.4798S} and \citet{HERA2023}. For consistency, the {\it Chandra} results have been converted to the same soft-band energy interval and initial mass function as the default \citet{Salpeter1955ApJ...121..161S} in {\it 21cmFAST}. Overall, constraints from 21-cm power spectra follow the trend of an enhanced X-ray luminosity towards higher redshits, which are typically expected for the lower-metallicity environment of early galaxies \citep{Fragos2013ApJ...764...41F}. However, unlike \citet{HERA2023}, the current MWA limits provide only a mild constraint, yielding $L_{\rm X<2keV}/{\rm SFR}>10^{39}{\rm erg\ s^{-1}\ M_{\odot}^{-1} yr}$ and still leaving local values permissible. This suggests that even modest X-ray emission can produce sufficient heating to explain the observed limits of the 21-cm power spectra. 


\section{Discussions and Future Work}
\label{sec:discussion}
The MWA results mark an important step in constraining the thermal state of the early IGM and the nature of heating during reionization. Future observations with increased sensitivity, such as those anticipated from the Square Kilometre Array (SKA), will tighten constraints on $L_{\rm X<2keV}/{\rm SFR}$ as well as other properties of the first galaxies \citep{Park2019MNRAS.484..933P,Qin2021MNRAS.501.4748Q}, enabling a more detailed understanding of the interplay between star formation, heating and the 21-cm signal. Combined with James Webb Space Telescope observations of early galaxies, these studies will refine our picture of galaxy formation and IGM evolution in the early Universe. 

Additionally, MWA results can be improved by refining our current data processing and power spectrum estimation tools. Below are some potential identified avenues:



\begin{itemize}
\item Van Vleck Correction: Addressing non-linearities in the correlator responses caused by signal loss during quantization (see \citealt{Benkevitch2016, Pyxie_thesis} for details).

\item Kernel Density Estimation: Reducing contamination within the EoR window by isolating non-Gaussian components introduced by foregrounds, using a kernel density estimator.

\item Refinement of RFI Flagging: While our current RFI mitigation techniques have successfully detected many instances of bright, or  rapidly changing RFI, there is significant room for improvement in this area. Other flagging algorithms currently under development, such as EAVILS, show great potential for detecting faint and slowly moving RFI.


\item Using auto-correlations for calibration: Applying bandpass solutions derived from auto-correlated visibilities can mitigate spurious spectral structures introduced into the calibration solutions by imperfections in sky and instrument models \citep{Barry2019a, Barry2019b, Li2019}. This method was first implemented in \citet{EwallWice2016} to reduce residual spectral structures imprinted on foregrounds by reflections in the signal chain.

\item {Re-performing direction-independent calibration: \citet{Shintaro2021} demonstrated that ionospheric correction is essential even during direction-independent calibration. This process involves recalibrating for direction-independent gains using an updated calibration model that incorporates ionospherically corrected sources.}
\end{itemize}

\section{Conclusion}
\label{sec:conclusion}
In this study, we presented the data processing and power spectrum estimation pipeline implemented using the Hyperdrive/CHIPS pipeline. Our approach builds upon the systematic mitigation strategy described in \citet{Nunhokee2024}, with several modifications designed to address the challenges posed by incorporating data from Phase II MWA configurations.

Starting with an initial dataset of 19,698 observations, our methodology identified 52$\%$ of the data as high quality and suitable for power spectrum analysis.  Observations from pointings at -3 were excluded due to the prominent aliasing effects of the Galactic plane. 

We constructed power spectra at three redshifts: $z=6.5$, $z=6.8$ and $z=7.0$. At $z=6.5$, the most stringent upper limit was determined to be $(30.2)^2$ mK$^2$ at $k=0.18, h$ Mpc$^{-1}$. At $z=6.8$, the upper limit was measured as $(31.3)^2$ mK$^2$ at $k=0.18, h$ Mpc$^{-1}$, while at $z=7.0$, it was evaluated as $(39.1)^2$ mK$^2$ at $k=0.21, h$ Mpc$^{-1}$.

Our results provide the first evidence of a heated IGM at redshfits $z=6.5$ to $z=7.0$. By ruling out models with excessive power, these limits establish a lower bound of $L_{\rm X<2keV}/{\rm SFR}>10^{39}{\rm erg\ s^{-1}\ M_{\odot}^{-1} yr}$ for star-forming galaxies, where the soft-band X-ray luminosity per unit star formation rate governs the heating of the early Universe.

We validated the results produced from Hyperdrive/CHIPS against an end-to-end simulation and an independent pipeline, FHD/$\epsilon$ppsilon. The simulation, performed using WODEN\footnote{\url{https://github.com/JLBLine/WODEN}}, derived a normalization factor that matched the observed power levels. Additionally, FHD/$\epsilon$ppsilon produced power spectra consistent with Hyperdrive/CHIPS at low $k$-modes when applied to a subset of the final integration. However, at high $k$-modes, Hyperdrive/CHIPS exhibited significantly higher power levels. These discrepancies are attributed to the 160~kHz averaging performed by FHD/$\epsilon$ppsilon. Beyond this, Hyperdrive/CHIPS does not display any unique bias. 


\section{Acknowledgements}
\label{sec:acknowledgments}
This research was supported by the Australian Research Council Centre of Excellence for All Sky Astrophysics in 3 Dimensions (ASTRO 3D), through project number CE170100013. This scientific work uses data obtained from \textit{Inyarrimanha Ilgari Bundara} / the Murchison Radio-astronomy Observatory. We acknowledge the Wajarri Yamaji People as the Traditional Owners and native title holders of the Observatory site. Establishment of CSIRO’s Murchison Radio-astronomy Observatory is an initiative of the Australian Government, with support from the Government of Western Australia and the Science and Industry Endowment Fund. Support for the operation of the MWA is provided by the Australian Government (NCRIS), under a contract to Curtin University administered by Astronomy Australia Limited. This work was supported by resources provided by the Pawsey Supercomputing Research Centre with funding from the Australian Government and the Government of Western Australia. This work was supported by resources awarded under Astronomy Australia Ltd’s merit allocation scheme on the OzSTAR national facility at Swinburne University of Technology. OzSTAR is funded by Swinburne University of Technology and the National Collaborative Research Infrastructure Strategy (NCRIS). The International Centre for Radio Astronomy Research (ICRAR) is a Joint Venture of Curtin University and The University of Western Australia, funded by the Western Australian State government. This research received technical support from the Australian SKA Regional Centre (AusSRC).
The inferences were performed on the Gadi supercomputer. YQ acknowledges HPC resources from the ANUMAS and support from the ARC Discovery Early Career Researcher Award (DECRA) through fellowship DE240101129. NB acknowledges support by the ARC Discovery Early Career Researcher Award (DECRA) through project number DE240101377. JCP acknowledges support from the U.S. National Science Foundation grant \#2106510. MFM, BH, and EL acknowledge support from the U.S. National Science Foundation grants \#2107538 and \#2228990.

\appendix
\label{apendix}
\begin{table*}
\caption{$2\sigma$ upper limits on the EoR power at $z=6.5, 6.8$ and $7.0$, denoted by $\Delta^2$ for East--West and North--South polarizations. The thermal noise $\Delta^2_{therm}$ and sample variance $\Delta^2_{sample}$ are listed alongside each polarization. The number of observations and $k$-cuts used are displayed in Table~\ref{table:results}.} 
\label{tab:allz_results}
\begin{tabular}{cc|ccc|ccc}
   \hline
   & & \multicolumn{3}{c|}{East--West (mK$^2$)} & \multicolumn{3}{c}{North--South (mK$^2$)} \\
   \hline
   $z$ &$k(h$ Mpc$^{-1})$ & $\Delta^2$ & $\Delta^2_{therm}$ & $\Delta^2_{sample}$ & $\Delta^2$ & $\Delta^2_{therm}$ & $\Delta^2_{sample}$ \\
   \hline
$6.5$ & $0.142$ & $(56.7)^2$ & $(2.7)^2$ & $(13.8)^2$ & $(57.9)^2$ & $(2.7)^2$ & $(14.1)^2$\\
   &  $0.177$ &  {$\bf ( 30.2)^2$} & $(3.7)^2$ & $(7.4)^2$ & $(39.5)^2$ & $(3.7)^2$   & $(9.6)^2$\\
   &  $0.212$ &  $(34.6)^2$   &  $(4.9)^2$ & $(8.4)^2$ & {\bf $ \bf (39.2)^2$} & $(4.8)^2$ & $(9.5)^2$\\
   &  $0.248$ &  $(96.1)^2$  &  $(6.2)^2$ & $(23.4)^2$ & $(87.7)^2$ & $(6.1)^2$ & $(21.3)^2$\\
   &  $0.283$ &  $(503.4)^2$  & $(8.9)^2$ & $(158.9)^2$ & $(487.0)^2$ &$(8.9)^2$ & $(153.7)^2$ \\
   &  $0.319$ &  $(416.6)^2$ &  $(10.7)^2$ & $(111.2)^2$ & $(388.2)^2$ & $(10.7)^2$ &  $(103.7)^2$\\
   &  $0.354$ &  $(2046.8)^2$ & $(10.5)^2$& $(498.6)^2$ & $(2001.1)^2$ & $(10.5)^2$ &  $(487.5)^2$\\
   \hline \hline 
$6.8$ & $0.142$ & $(77.4)^2$ & $(2.9)^2$ & $(18.5)^2$ & $(115.4)^2$ & $(2.9)^2$ & $(27.6)^2$\\
   &  $0.177$ &  {$\bf (31.3)^2$} & $(4.1)^2$ & $(7.5)^2$ & {$\bf (79.3)^2$} & $(4.1)^2$  & $(18.9)^2$\\
   &  $0.212$ &  $(36.6)^2$   &  $(5.3)^2$ & $(8.7)^2$ & $(96.0)^2$ & $(5.3)^2$ & $(22.9)^2$\\
   &  $0.248$ &  $(110.1)^2$  &  $(6.7)^2$ & $(26.3)^2$ & $(157.7)^2$ & $(6.7)^2$ & $(37.7)^2$\\
   &  $0.283$ &  $(424.7)^2$  & $(8.2)^2$ & $(101.5)^2$ & $(460.3)^2$ &$(8.2)^2$ & $(110.0)^2$ \\
   &  $0.319$ &  $(1660.7)^2$ &  $(9.8)^2$ & $(396.9)^2$ & $(1759.0)^2$ & $(9.7)^2$ &  $(420.4)^2$\\
   &  $0.354$ &  $(5189.7)^2$ & $(11.5)^2$& $(1240.4)^2$ & $(5307.7)^2$ & $(11.5)^2$ &  $(1268.5)^2$\\
   \hline \hline 
   $7.0$ & $0.142$ & $(91.7)^2$ & $(3.3)^2$ & $(22.5)^2$ & $(100.8)^2$ & $(3.3)^2$ & $(24.8)^2$\\
   &  $0.177$ &  $(40.3)^2$ & $(4.6)^2$ & $(9.9)^2$ & {$\bf (55.3)^2$} & $(4.6)^2$   & $(13.6)^2$\\
   &  $0.212$ &  { $\bf (39.1)^2$}   &  $(6.1)^2$ & $(9.6)^2$ & $(60.8)^2$ & $(6.1)^2$ & $(14.9)^2$\\
   &  $0.248$ &  $(123.3)^2$  &  $(7.6)^2$ & $(30.3)^2$ & $(165.1)^2$ & $(7.6)^2$ & $(40.6)^2$\\
   &  $0.283$ &  $(493.4)^2$  & $(9.3)^2$ & $(121.3)^2$ & $(556.8)^2$ &$(9.3)^2$ & $(136.9)^2$ \\
   &  $0.319$ &  $(1691.3)^2$ &  $(11.1)^2$ & $(415.8)^2$ & $(1935.2)^2$ & $(11.1)^2$ &  $(475.7)^2$\\
   &  $0.354$ &  $(4932.2)^2$ & $(13.1)^2$& $(1212.4)^2$ & $(5542.4)^2$ & $(13.0)^2$ &  $(1362.4)^2$\\
   \hline \hline \\
   
\end{tabular}
\end{table*}

\bibliography{bibliography}{}
\bibliographystyle{aasjournal}

\end{document}